\documentclass[12pt,a4paper,fleqn]{article}
\usepackage[fleqn]{amsmath}
\usepackage{amsmath}
\usepackage{mathrsfs}
\usepackage{amssymb}
\usepackage{latexsym}
\usepackage{makeidx}
\usepackage{CJK}
\usepackage{indentfirst}
\usepackage{tikz}
\usepackage{array}
\usepackage{graphicx}
\usepackage{epsfig}
\usepackage{subfigure}
\usepackage{color}
\usepackage{cite}
\usepackage{enumerate}
\usepackage[percent]{overpic}
\usepackage{overpic}
\usepackage[top=0.5in, bottom=1in, left=0.5in, right=0.5in]{geometry}
\usepackage[colorlinks, linkcolor=blue, anchorcolor=blue, citecolor=blue]{hyperref}
\date{}
\title{\Large \textbf{Adjoint SU(5) GUT model with Modular $S_4$ Symmetry}}
\author{
Ya Zhao$^{1, }$  \footnote{\emph{E-mail address}: zhaoya@mail.ustc.edu.cn} ,
Hong-Hao Zhang$^{1, }$ \footnote{\emph{E-mail address}: zhh98@mail.sysu.edu.cn}\\
\bigskip
\\
{ $^{1}$ {\footnotesize
 \it School of Physics, Sun Yat-Sen University, GuangZhou, GuangDong 510275,
 P. R. China.}}}
\begin{document}
\normalsize
\maketitle
\begin{abstract}
We study the textures of SM fermion mass matrices and their mixings in a supersymmetric
adjoint SU(5) Grand Unified Theory with modular $S_4$ being the horizontal symmetry.
The Yukawa entries of both quarks and leptons are expressed by modular forms with lower
weights. Neutrino sector has an adjoint SU(5) representation 24 as matter superfield,
which is a triplet of $S_4$. The effective light neutrino masses is generated
through Type-III and Type-I seesaw mechanism. The only common complex parameter in
both charged fermion and neutrino sectors is modulus $\tau$. Down-type quarks
and charged leptons have the same joint effective operators with adjoint scalar
in them, and their mass discrepancy in the same generation depends on Clebsch-Gordan
factor. Especially for the first two generations the respective Clebsch-Gordan
factors made the double Yukawa ratio $y_dy_{\mu}/y_ey_s=12$, in excellent agreement with
the experimental result. We reproduce  proper CKM mixing parameters and all nine
Yukawa eigenvalues of quarks and charged leptons. Neutrino masses and
MNS parameters are also produced properly with normal ordering is preferred.
\end{abstract}
%%%%%%%%%%%%%%%%%%%%%%%%%%%%%%%%%%%%%%%%%%%%%%%%%%
\thispagestyle{empty}
\newpage
%\begin{multicols}{2}
\section{Introduction}\label{S1}
Despite of great success, especially the discovery of
Higgs boson~\cite{Aad/2012ATLAS,Chatrchyan/2012CMS}, the standard model (SM) still has
some problems unsolved in particle physics. One of the problems is the origin of flavor
structure for the SM fermions, which mainly refers to the enormous mass difference among
different generations, and distinct mixing patterns between lepton and quark sector.

The masses of charged fermions, from lightest electron of MeV level to heaviest top
quark being 173GeV, span almost 5 orders of magnitude. The situation becomes even
worse when the neutrino sector is contained, since the neutrino masses of sub-eV
are nearly 7 orders of magnitude smaller than that of electron. Besides the absolute
mass scale and mass ordering are still yet to be determined by future high-precision
neutrino experiments. Apart from the large mass hierarchy, quark CKM parameters manifest
a mixing pattern of small angles~\cite{Tanabashi/2018PDG}.
However the situation is completely different in
lepton MNS mixing matrix. The precise neutrino oscillation data has provided
us a picture of two large angles $\theta_{12}$ and $\theta_{23}$, and one small
angle $\theta_{13}$, which is comparable with the quark Cabibbo angle $\theta_C$.

It is an unsolved puzzle to interpret the observed flavor structure in quark and lepton
sector. The flavor parameters, include all the fermion masses, mixing angles as well as
CP violating phases arise from the dependence of Yukawa on flavor. Since symmetry
plays an important role in physics, it is worth to constrain the Yukawa interactions
with the supervision of flavor symmetry, hence the mass hierarchies and mixing patterns
of SM fermions could be interpreted by the additional symmetry beyond the gauge symmetry
of the SM. But note that flavor symmetry is not the unique top-down scenario
to understand the flavor structure. The other possible top-down approaches include,
such as anarchy~\cite{Hall/2000ANX},
extra dimension~\cite{Dienes/1999XDS,Arkani-Hamed/2002NXD}
and string theory~\cite{Blumenhagen/2007DBR,Ibanez/2007STN,Antusch/2007STN}.

For the last two decades, the precise measurement of flavor parameters, especially
that of lepton mixing angles, has motivated the model building by using discrete symmetry
group to elucidate the different flavor structures between quarks and leptons. Non-Abelian
discrete groups, such as lower order groups $A_4$, $A_5$, $S_4$ and other ones with higher
order, are wildly used in such kind of works. And it is indeed easy to reproduce at least
the two large leptonic mixing angles. For reviews see Refs.
\cite{Altarelli/2010RMP,Ishimori/2010NAD,King/2013RPP,King/2015REV,King/2017UNM,
Meloni/2017FMN}. In the models apart from the essential Higgs, some extra scalar
fields called flavons are introduced. They are singlets under SM gauge group but have
nontrivial representations under flavor group, whose vacuum orientation in flavor
space can induce specific Yukawa textures for fermions. In neutrino sector the famous
Tri-Bimaximal mixing~\cite{Harrison/2002HPS,Harrison/2002TBM} is ubiquitous in
many realistic models.

Non-Abelian discrete groups as flavor symmetries are success in explaining the
leptonic mixing pattern with large angles, with the price of introducing a number
of gauge singlets called flavon fields in conventional studies. The flavor
symmetry is broken when the scalar flavons acquiring vacuum expectation values
(VEV) by non-trivial dynamical way. Such VEVs with specific configurations control
the flavor textures of fermions and thus a few of free coupling parameters appear
in the Yukawa entries. Besides the values of VEVs themselves are determined by
some other parameters. The more parameters in the traditional flavor models,
the less predictive power they have. And the fermion masses are still often acquired
by tuning the coupling parameters, even though the VEVs of the flavons can be nearly
fixed such that all the couplings can be naturally of the same order, e.g., of order one.

Recently the modular symmetry provides another possibility to interpret the flavor
issues~\cite{Feruglio/2017NMF}. The simplest modular symmetry implementation only
demands one complex field, called modulus $\tau$, as the unique source of modular
symmetry breaking when it develops a VEV, hence the vacuum configuration problem
is greatly simplified. In such a simplest framework the Yukawa couplings are just
modular forms which are the holomorphic functions of modulus $\tau$, then the
flavon fields are not indispensable ingredient for model building, and thus tremendously
reduce the amount of particle content and the free parameters of the theory.

The modular invariance models are based on the level $N$ finite modular group $\Gamma_N$,
e.g., from $N=2$ to 5, $\Gamma_2\simeq S_3$
~\cite{Kobayashi/2018NFM,Okada/2019MST,Kobayashi/2020SRG,Mishra/2020MLG},
$\Gamma_3\simeq A_4$
~\cite{Feruglio/2017NMF,Criado/2018MDN,Okada/2019CPV,
Okada/2020CTA,Ding/2019MAF,Kobayashi/2020ASM,Nomura/2019TLP,
Kobayashi/2019BNV,Nomura/2019MAF,Anda/2020GAF,KingSJ/2020FMH,Asaka/2020MAF,Behera/2020MAL},
$\Gamma_4\simeq S_4$
~\cite{Penedo/2018LMS,Novichkov/2018MSL,MedeirosVarzielas/2019MPS,Ding/2019MAS,
Kobayashi/2019ASM,King/2019DMS,Criado/2019LQP,WangX/2020QMS,WangX/2020DRS}
and $\Gamma_5\simeq A_5$
~\cite{Novichkov/2018AFM,Ding/2019AFM,Criado/2019LQP},
which are all inhomogeneous modular groups. In such models the Yukawa
couplings are weight $k$ modular forms with $k$ being even numbers. Most of the studies
focus on the lepton flavor issues, while few of them include quark sector
~\cite{Okada/2019CPV,Okada/2019MST,KingSJ/2020FMH,Okada/2020CTA}. The unified
quark and lepton models can be implemented in the context of SU(5) grand
unified theories (GUT) combined with the modular symmetries
\cite{Anda/2020GAF,Kobayashi/2020SRG}. Besides the topics on the
fermion flavor structures, the related phenomenological issues has been discussed in
those works, such as the dark matter models~\cite{Nomura/2019MAF}, baryon number
violation~\cite{Kobayashi/2019BNV} and leptogenesis
~\cite{Asaka/2020MAF,Behera/2020MAL,Mishra/2020MLG}.
On the other side, the double covering of finite modular group which is homogeneous,
has been used as flavor symmetry as well
~\cite{LiuXG/2019DCA,LuJN/2020TXZ,Novichkov/2020DCS,LiuXG/2020DCF,
WangX/2020DCF,YaoCY/2020DCF}.

Motivated by the phenomenological viable mass ratios between quarks and leptons,
and the idea of Yukawa couplings can be modular forms, in the study we combine the modular
flavor symmetry $\Gamma_4\simeq S_4$ with the supersymmetric
adjoint SU(5) GUT~\cite{Perez/2007SAS} to forge the flavor textures of quarks
and leptons simultaneously. In refs.~\cite{Anda/2020GAF,Kobayashi/2020SRG} the neutrino
masses are generated via type-I
~\cite{Minkowski/1977SS,Yanagida/1979SS,Mohapatra/1980SS,Gell-Mann/1979sug}
seesaw by adding at least two gauge singlets, i.e.,
right-handed heavy neutrinos. However there exist another two possibilities in SU(5) GUT
to produce light neutrino masses: first is type-II seesaw mechanism
~\cite{Lazarides/1981SGT,Schechter/1980NUM,Mohapatra/1981NMV} by adding
an extra Higgs $\textbf{15}_H$, and second is Type-I plus
Type-III~\cite{Foot/1989SS3} seesaw by adding fermionic field in the \textbf{24}
dimensional representation. The two cases in SU(5) GUT have not been explored
yet in the recent modular flavor models. In this paper we explore for the first
time the second scheme to produce effective light neutrino masses. Meanwhile
we would like to give rise to a realistic Yukawa ratios between charged leptons
and down-quarks. The Yukawa ratio in each generation is directly derived from the
novel Clebsch-Gordan (CG) factor, which can be comparable to the phenomenological
values at GUT scale. For the modular flavor model built on SU(5) GUT, the
flavon-free model is such that the double Yukawa eigenvalue ratio
$y_{\mu}y_d/y_ey_s$ equal to 12 for the first and second families of charged leptons
and down quarks. The other viable cases which can generate the acceptable ratios,
however, still require the flavon or so called weighton to compensate the loss of
mass dimension.

There is no flavon but a modulus $\tau$ in the modular symmetry breaking sector,
meanwhile the gauge symmetry is broken by one adjoint scalar $H_{24}$. We aim
at minimizing the amount of free coupling parameters in the unified model.
For higher weight modular forms, it will bring more free coupling parameters,
and thus decrease the predictive power of the model. Thus we use lower weight
modular forms to give the modular invariant operators.

In order to achieve the above goal, we should strictly contrain the representations
and weights under $S_4$ for the fields. To be specific, all the \textbf{10}-dimensional
matter superfields are $S_4$ singlets and have distinct weights. The three families
of $\bar{\textbf{5}}$s are divided into an doublet and a singlet but have the same weight.
At last the \textbf{24} fermionic superfields are sealed in a triplet. For the scalar
sector, the $\textbf{5}_H$, $\textbf{45}_H$ and $\textbf{24}_H$ are just singlets with
distinct weights. Please see Table.~\ref{ta:tab11} for details.

In quark and charged lepton sector the Yukawa matrices are very sparse with
several texture zeros. The adjoint scalar $H_{24}$ couples to the matter superfields
in down quark sector, then it induces the novel ratio of CG factors 1/2 and 6 for
the first two families of leptons and down quarks, and 3/2 for the third one.
In neutrino sector we introduce the \textbf{24} dimensional matter field rather
than a gauge singlet to produce the neutrino masses through Type-I and Type-III
seesaw mechanism. Since we introduce the adjoint matter fields rather than gauge
singlets to produce the effective masses of light neutrinos, the Yukawa matrices
are slightly different for the heavy $\rho_3$ and $\rho_0$ in the \textbf{24}.
And the same for Majorana mass matrices, since the mass terms include two
nontrivial couplings: the pure mass term which is the same for $\rho_3$ and $\rho_0$,
and the new interaction between \textbf{24} and $H_{24}$, which splits the masses
of $\rho_3$ and $\rho_0$. Such new interaction is of course absent in the models
which gauge singlets are responsible for seesaw mechanism.

The layout of the paper is arranged as follows. In Sec.~\ref{sec:Frame} we introduce
the framework for the model building work, including the brief review
on modular group, especially the one with level $N=4$, $\Gamma_4\simeq S_4$ and
the modular forms of weight 2 and 4. Then we give the basic aspects of adjoint SU(5)
GUT. In Sec.~\ref{sec:Model} we present the flavor model based on adjoint SU(5) GUT
combined with modular $S_4$ flavor symmetry. We show that the GUT flavor model can
be built without introducing a gauge singlet scalar. The entries of Yukawa matrices
of quarks and charged leptons have only modular forms in them.
In Sec.~\ref{sec:Phenomenology}, we first give the convention for Yukawa matrices and the
SUSY threshold corrections, then we present the data to be used
and perform the numerical fit to the Yukawa matrices with threshold effects included.
Sec.~\ref{sec:Summary} devotes to the summary of the study.

%%%%%%%%%%%%%%%%%%%%%%%%%%%%%%%%%%%%%%%%%%%%%%%%%%

\section{The Framework}\label{sec:Frame}
In the section we shall briefly discuss the framework and environment used for the
construction of model. First we give a brief review on the basic concepts of modular
symmetry with lower level $N$ and modular forms of weight $k$. The finite
modular group $\Gamma_4$ is used for our model building work, so we give the modular
forms with weight $k=2,4$. Secondly we introduce the basic aspects of adjoint SU(5)
grand unified theory, including the matter multiplets and scalar Higgs together with
the vacuum configurations of the scalars.

\subsection{Modular group and Modular forms}
The modular group $\overline{\Gamma}$ implies the linear fractional transformations
$\gamma$ that act on the complex $\tau$ in the upper-half complex plane
\begin{equation}\label{eq:modular}
  \tau\rightarrow \gamma\tau=\frac{a\tau+b}{c\tau+d},\quad a,b,c,d\in \mathbb{Z},
  \quad ad-bc=1,\quad \textrm{Im}\tau >0,
\end{equation}
The generators $S$ and $T$ are the two transformations satisfying
\begin{equation}\label{eq:ST}
  S^2=(ST)^3=\mathbb{I},
\end{equation}
with the representation matrices as
\begin{equation}\label{eq:ST matrix}
  S=\left(
      \begin{array}{cc}
        0 & 1 \\
        -1 & 0 \\
      \end{array}
    \right),\quad
  T=\left(
      \begin{array}{cc}
        1 & 1 \\
        0 & 1 \\
      \end{array}
    \right),
\end{equation}
lead to
\begin{equation}\label{eq:STtau}
  S: \tau\rightarrow -\frac{1}{\tau},\quad T: \tau\rightarrow \tau+1.
\end{equation}

We introduce the series of infinite normal subgroups $\Gamma(N)$, $N=1,2,3,\cdots$
of $SL(2,\mathbb{Z})$
\begin{equation}\label{eq:GammaN}
  \Gamma(N)=\Bigg\{
  \left(
  \begin{array}{cc}
  a & b \\
  c & d \\
  \end{array}
  \right)\in SL(2,\mathbb{Z}),\quad
  \left(
  \begin{array}{cc}
  a & b \\
  c & d \\
  \end{array}
  \right)=
  \left(
  \begin{array}{cc}
  1 & 0 \\
  0 & 1 \\
  \end{array}
  \right)(\textrm{mod}\;N)
  \Bigg\}.
\end{equation}
For $N=1,2$ we define $\overline{\Gamma}(N)=\Gamma(N)/\{\mathbb{I},-\mathbb{I}\}$
and for $N>2$ we have $\overline{\Gamma}(N)=\Gamma(N)$. Taking the quotient
$\Gamma_N\equiv\overline{\Gamma}/\overline{\Gamma}(N)$, one can obtain a finite subgroup
called finite modular group. Especially for $N\leq5$ the groups $\Gamma_N$ are isomorphic
to the permutation groups $\Gamma_2\simeq S_3, \Gamma_3\simeq A_4, \Gamma_4\simeq S_4$
and $\Gamma_5\simeq A_5$, which are ubiquitous in the construction of flavor models.

Modular forms $f(\tau)$ of weight $k$ and level $N$ are holomorphic functions transforming
under the groups $\Gamma(N)$ in the way as
\begin{equation}\label{eq:MF}
 f(\gamma\tau)=(c\tau+d)^kf(\tau),\quad \gamma\in\Gamma(N).
\end{equation}
Here $k$ is even number, and $N$ is natural. Given the weight $k$ and level $N$, the
modular forms span a linear space of dimension equals to $k+1$. The basis in the
linear space can be chosed such that the modular form $f_i(\tau)$ in a multiplet
transforms according to a unitary representation $\rho$ of the group $\Gamma_N$:
\begin{equation}\label{eq:MFrep}
  f_i(\gamma\tau)=(c\tau+d)^k\rho(\gamma)_{ij}f_j(\tau),\qquad \gamma\in\Gamma_N.
\end{equation}
In the study we take $N=4$ as the case of interest and construct an explicit grand
unified flavor model to elucidate the fermion masses and mixings. In the case of
lowest weight 2, there are 5 linear independent modular forms. These modular forms
are explicitly expressed in terms of Dedekind eta-function $\eta(\tau)$:
\begin{equation}\label{eq:etafunc}
  \eta(\tau)=q^{1/24}\prod^{\infty}_{n=1}(1-q^n),\qquad q=e^{i2\pi\tau}.
\end{equation}
To be specific, the modular forms are defined as the functions of $\eta(\tau)$ and
its derivatives of the form
\begin{equation}\label{eq:MFY}
  Y(c_1,\cdots,c_6|\tau)\equiv
   c_1\frac{\eta'(\tau+\frac{1}{2})}{\eta(\tau+\frac{1}{2})}
  +c_2\frac{\eta'(4\tau)}{\eta(4\tau)}
  +c_3\frac{\eta'(\frac{\tau}{4})}{\eta(\frac{\tau}{4})}
  +c_4\frac{\eta'(\frac{\tau+1}{4})}{\eta(\frac{\tau+1}{4})}
  +c_5\frac{\eta'(\frac{\tau+2}{4})}{\eta(\frac{\tau+2}{4})}
  +c_6\frac{\eta'(\frac{\tau+3}{4})}{\eta(\frac{\tau+3}{4})},
\end{equation}
with the coefficients $c_1+\cdots+c_6=0$. For the case of weight 2, the basis is
comprised of five modular forms as follow:
\begin{eqnarray}\label{eq:MFY5}
  & & Y_1(\tau)\equiv Y(1,1,\omega,\omega^2,\omega,\omega^2|\tau),\\
  & & Y_2(\tau)\equiv Y(1,1,\omega^2,\omega,\omega^2,\omega|\tau),\\
  & & Y_3(\tau)\equiv Y(1,-1,-1,-1,1,1|\tau),\\
  & & Y_4(\tau)\equiv Y(1,-1,-\omega^2,-\omega,\omega^2,\omega|\tau),\\
  & & Y_5(\tau)\equiv Y(1,-1,-\omega,-\omega^2,\omega,\omega^2|\tau),
\end{eqnarray}
with $\omega\equiv e^{2\pi i/3}$. The modular forms $(Y_1,Y_2)^T$ and $(Y_3,Y_4,Y_5)^T$
transform as an doublet and triplet of $S_4$ respectively, i.e., they are denoted as
\begin{equation}\label{eq:Yw2}
  Y_{\bf{2}}(\tau)\equiv
  \left(
  \begin{array}{c}
  Y_1(\tau) \\
  Y_2(\tau) \\
  \end{array}
  \right),\quad
  Y_{\bf{3}'}(\tau)\equiv
  \left(
  \begin{array}{c}
  Y_3(\tau) \\
  Y_4(\tau) \\
  Y_5(\tau) \\
  \end{array}
  \right).
\end{equation}
One can get modular forms of higher weights from the tensor productions of weight 2
modular forms, and thus obtain different irreps of $S_4$. Taking weight $k=4$ for example,
we then have 9 independent modular forms, and they are arranged into the following
singlets and multiplets irreps of $S_4$
\begin{eqnarray}\label{eq:Yw4}
& &  Y^{(4)}_{\bf{1}}=Y_1Y_2,\quad
  Y^{(4)}_{\bf{2}}=
  \left(
  \begin{array}{c}
  Y^2_2 \\
  Y^2_1 \\
  \end{array}
  \right),\nonumber\\
& &  Y^{(4)}_{\bf{3}}=
  \left(
    \begin{array}{c}
      Y_1Y_4-Y_2Y_5 \\
      Y_1Y_5-Y_2Y_3 \\
      Y_1Y_3-Y_2Y_4 \\
    \end{array}
  \right),\quad
    Y^{(4)}_{\bf{3}'}=
  \left(
    \begin{array}{c}
      Y_1Y_4+Y_2Y_5 \\
      Y_1Y_5+Y_2Y_3 \\
      Y_1Y_3+Y_2Y_4 \\
    \end{array}
  \right).
\end{eqnarray}
We use the above modular forms of weight 2 and 4 to construct our modular invariant
model in Sec.~\ref{sec:Model}. For higher weight modular forms,
one may refer to Ref.~\cite{Novichkov/2018MSL} for further information.

\subsection{Basic aspects of Adjoint SU(5)}
In this part we shall give the fermion sector and the scalars used for the model setup.
According to the distinct representations under the gauge symmetry,
the matter fields are divided into the following three parts: anti-fundamental
$\bar{\textbf{5}}$, anti-symmetrical tensor \textbf{10} and the
adjoint \textbf{24}. We assume the usual three generations of $\bar{\textbf{5}}$ and
\textbf{10}, which have the decomposition under the SM gauge group
$SU(3)_C\otimes SU(2)_L\otimes U(1)_Y$. The theory of renormalizable adjoint SU(5) 
can be seen in ref.~\cite{Perez/2016RAD}. Concerning our model, we promote the matter
fields to superfields in Minimal Supersymmetry Standard Model (MSSM) and embed them into
SU(5) representations $\bar{\textbf{5}}$ and \textbf{10}. For $\bar{\textbf{5}}$ we denote
\begin{equation}\label{eq:F5}
\qquad  F_i = (d^{c}_{R} \quad d^{c}_{B} \quad d^{c}_{G} \quad e \quad -\nu)_i
            = d^c_i\oplus\ell_i
\end{equation}
and for \textbf{10}
\begin{equation}\label{eq:T10}
\qquad  T_{i}=\frac{1}{\sqrt{2}}\left(
                            \begin{array}{ccccc}
                              0 & -u^{c}_{G} & u^{c}_{B} & -u_{R} & -d_{R} \\
                              u^{c}_{G} & 0 & -u^{c}_{R} & -u_{B} & -d_{B} \\
                              -u^{c}_{B} & u^{c}_{R} & 0 & -u_{G} & -d_{G} \\
                              u_{R} & u_{B} & u_{G} & 0 & -e^{c} \\
                              d_{R} & d_{B} & d_{G} & e^{c} & 0 \\
                            \end{array}
                          \right)_{i}
     =u^c_i\oplus q_i\oplus e^c_i,
\end{equation}
where $i=1,2,3$ indicates the family indices of standard model (SM),
and $R, B, G$ stand for the color indices. The adjoint matter field \textbf{24} reads
\begin{equation}\label{eq:24A}
  \textbf{24}=\frac{1}{\sqrt{2}}
  \left(
  \begin{array}{cc}
  \frac{1}{\sqrt{2}}\boldsymbol{\lambda}.\boldsymbol{\rho}_8-\sqrt{\frac{2}{15}}\rho_0
  & \rho_{(3,2)} \\
  \rho_{(\bar{3},2)}
  & \frac{1}{\sqrt{2}}\boldsymbol{\sigma}.\boldsymbol{\rho}_3+\sqrt{\frac{3}{10}}\rho_0 \\
  \end{array}
  \right),
\end{equation}
in which $\lambda^i$ are the Gell-Mann matrices. The scalar fields contain the
following Higgs fields
\begin{eqnarray}\label{eq:Higgs}
  & & \textbf{5}_H=T_{(3,1)}\oplus H_1, \nonumber\\
  & & \textbf{45}_H=S_8\oplus S_{(\bar{6},1)}\oplus S_{(3,3)}
  \oplus S_{(\bar{3},2)}\oplus S_{(3,1)} \oplus S_{(\bar{3},1)}\oplus H_2,\nonumber\\
  & & \textbf{24}_H=\Sigma_8 \oplus \Sigma_3 \oplus \Sigma_{(3,2)}
  \oplus \Sigma_{(\bar{3},2)} \oplus \Sigma_{24}.
\end{eqnarray}
The VEVs of the scalars are listed as the follwing form
\begin{eqnarray}\label{eq:HiggsVEV}
  & & \langle\textbf{5}_H\rangle=\frac{v_5}{\sqrt{2}}(0,0,0,0,1)^T,\label{eq:H5VEV}\\
  & & \langle\textbf{24}_H\rangle=v_{24}\textrm{diag}(1,1,1,-3/2,-3/2),\label{eq:H24VEV}\\
  & & \langle\textbf{45}_H\rangle^{i5}_j
  =\frac{v_{45}}{\sqrt{2}}[\textrm{diag}(1,1,1,-3,0)]^i_j,\quad
  \langle\textbf{45}_H\rangle^{in}_j=0,\label{eq:H45VEV}
\end{eqnarray}
where $i,j=1,...,5$, $n=1,...,4$ and $v=\sqrt{|v_5|^2+24|v_{45}|^2}=246\textrm{GeV}$.

\section{Adjoint SU(5) GUT Flavor Model}\label{sec:Model}
Now let us present the supersymmetric adjoint SU(5) GUT flavor model in details. First we
assign the representations and weights of chiral supermultiplets. For sake of simplicity
we shall impose certain constraints on the assignments. For matter fields, three
generations of $\bf{10}$ dimensional representations, denoted by $T_{1,2,3}$, are
all $S_4$ singlets, i.e., $\rho_{T_i}\sim \bf{1}$, $(i=1,2,3)$, but have distinct
weights for each family. The first two generations of $\bf{\bar{5}}$ dimensional
representations, named as $F_{1,2}$, are assigned to be a doublet of $S_4$, i.e.,
$F=(F_1,F_2)$, while the third generation $F_3$ is a $S_4$ singlet. Nevertheless
the three $\bf{\bar{5}}$s have the same weight. In addition to the \textbf{10}-
and $\bar{\textbf{5}}$-dimensional representations in SU(5) GUTs, the adjoint
$\bf{24}$ dimensional matter superfields $A$ have also three families, and
they are assigned to be an triplet of $S_4$, denoted by $A=(A_1,A_2,A_3)$.

For the Higgs sector $H_5$, $H_{45}$, $H_{\bar{5}}$ and $H_{\overline{45}}$
are responsible for the electroweak symmetry broken, while the GUT gauge group is
broken by an adjoint scalar field $H_{24}$. We shall not intend to explain the
mass hierarchy of charged fermions, which is beyond the power of modular symmetry.
And in order to made the coupling terms minimality, we do not introduce any flavon
fields like conventional models did.  Instead all the Yukawa couplings are modular
forms with specific weights to ensure the modular invariance. The only modular symmetry
breaking source arises from the modulus $\tau$ developing its VEV. For simplicity
the modular weights are assigned such that the Yukawa couplings have as lower weights
as possible. The representations and weights of the chiral supermultiplets are listed
in the Table~\ref{ta:tab11}.
\begin{table}[!htbp]%\footnotesize
\renewcommand\arraystretch{1.2}
  \centering
  \caption{Field content and their representation assignments under the gauge
  group SU(5), modular $S_4$ and their weights $k_I$ in the model.}\label{ta:tab11}
\begin{tabular}{|c|ccccccccccc|}
  \hline
  Fields  & $T_1$ & $T_2$ & $T_3$ & $F=(F_1,F_2)$ & $F_3$ & $A$
  & $H_5$ & $H_{\bar{5}}$ & $H_{45}$ & $H_{\overline{45}}$ & $H_{24}$ \\
  \hline
  $SU(5)$ & $\bf{10}$ & $\bf{10}$ & $\bf{10}$ & $\bf{\bar{5}}$ & $\bf{\bar{5}}$
  & $\bf{24}$ & $\bf{5}$ & $\bf{\bar{5}}$ & $\bf{45}$
  & $\overline{\bf{45}}$ & $\bf{24}$ \\
  $\Gamma_4\equiv S_4$ & $\bf{1}$ & $\bf{1}$ & $\bf{1}$ & $\bf{2}$ & $\bf{1}$
  & $\bf{3}'$ & $\bf{1}$ & $\bf{1}'$ & $\bf{1}$ & $\bf{1}$ & $\bf{1}'$ \\
  $k_I$ & 0 & $-1$ & 1 & 1 & 1 & $-1$ & $-2$ & 1 & 0 & $-1$ & $-2$ \\
  \hline
\end{tabular}
\end{table}

\subsection{SU(5) breaking}
For the purpose of our model construction, we start with SU(5) breaking superpotential.
Since we have only one adjoint scalar field $H_{24}$ in the model, the gauge symmetry
broken into SM gauge group is realized by the adjoint scalar acquiring the vacuum
expectation value of $H_{24}$. The SU(5) breaking superpotential is simply as
\begin{equation}\label{eq:WH24}
  \mathcal{W}_{24}=M_{24}Y^{(4)}_{\textbf{1}}\textrm{Tr}H^2_{24}
  +\lambda Y^{(6)}_{\textbf{1}'}\textrm{Tr}H^3_{24},
\end{equation}
then one can get the VEV of $H_{24}$, i.e., $\langle H_{24}\rangle$
in eq.~\eqref{eq:H24VEV} with
\begin{equation}\label{eq:v24}
  v_{24}=\frac{4M_{24}Y^{(4)}_{\textbf{1}}}{3\lambda_1Y^{(6)}_{\textbf{1}'}}.
\end{equation}
Besides causing the GUT breaking, the adjoint scalar field $H_{24}$ would also couple
to the matter fields leading to novel Clebsch-Gordan factors, and especially the
exotic Yukawa coupling ratios between down quark sector and charged lepton sector at
GUT scale. See the next section for details.

\subsection{Charged Fermions}\label{subsec:chargefermion}
In this section we will present the Yukawa couplings for charged fermions. Since the
flavons are not an essential part of the model, the relevant Yukawa coupling terms in
up-type quark sector manifest in renormalisable level, while those in down-type quarks
and charged lepton sector show in effective operators.
Since the quarks have enormous mass hierarchies but small mixing angles, the resulting
Yukawa matrices should be controllable in the entries. A practical viable scheme is
to induce texture zeros for Yukawa matrices. And another keypoint is the GUT-scale
Yukawa ratios between leptons and down-quarks have to fulfil certain phenomenological
constraints from experiments as well.

\begin{figure}[htb]
  \centering
  {
  \label{fig:TTH} %% label for first subfigure
  \includegraphics[width=3.0in]%[bb=0 0 260 175,width=3.2in]
  {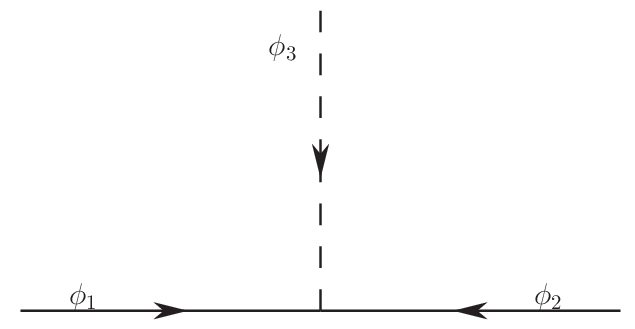}}
\caption{\small The supergraph for dimension 4 operators.}
\end{figure}

The Yukawa coupling terms for up quarks in superpotential involve two Higgs fields,
i.e., $H_5$ and $H_{45}$. Because of the symmetry constraints only the
following couplings are allowed
\begin{equation}\label{eq:wupII}
  \mathcal{W}_u=\alpha_uT_1T_1H_{45}+\beta_uT_2T_2H_5Y^{(4)}_{\bf{1}}
  +\gamma_uT_3T_3H_5+g_uT_2T_3H_{45},
\end{equation}
which leads to a block diagonal mass matrix
\begin{equation}\label{eq:massupII}
  M_u=\left[
  \begin{array}{ccc}
  \alpha_uv_{45} & 0 & 0 \\
  0 & \beta_uY_1Y_2v_5 & g_uv_{45} \\
  0 & -g_uv_{45} & \gamma_uv_5 \\
  \end{array}
  \right].
\end{equation}
The operators of dimension 4, such as those in Eq.~\eqref{eq:wupII}, are expressed by
the diagram in Figure.~\hyperref[fig:TTH]{1}
~\footnote{The supergraphs were drawn with \texttt{JaxoDraw}
~\cite{Binosi/2004JXD,Binosi/2009JXD}.},
in which the dashed line stands for scalar
Higgs and solid lines are matter multiplets. In Table~\ref{ta:dim4} we give the
list of dimension 4 operators corresponding to Figure.~\hyperref[fig:TTH]{1}
including those in the next sections.

\begin{table}[!htbp]\footnotesize
\renewcommand\arraystretch{1.2}
  \centering
\begin{tabular}{|c|cccccccc|}
  \hline
  \hline
 Field & 1 & 2 & 3 & 4 & 5 & 6 & 7 & 8 \\
  \hline
  $\phi_1$ & $T_1$ & $T_2$ & $T_3$ & $T_2$ & $T_1$ & F & $F_3$ & A\\
  $\phi_2$ & $T_1$ & $T_2$ & $T_3$ & $T_3$ & $F_3$ & A & A & A\\
  $\phi_3$ & $H_{45}$ & $H_5$ & $H_5$ & $H_{45}$ & $H_{\overline{45}}$
  & $H_5$ & $H_5$ & $H_{24}$\\
  \hline
  \hline
\end{tabular}
\caption{\small The list for dimension 4 operators corresponding
  to Figure.~\hyperref[fig:TTH]{1}.
  }\label{ta:dim4}
\end{table}

\begin{figure}[htb]
  \centering
  {
  \label{fig:TFHH} %% label for first subfigure
  \includegraphics[width=4.0in]%[bb=0 0 260 175,width=3.2in]
  {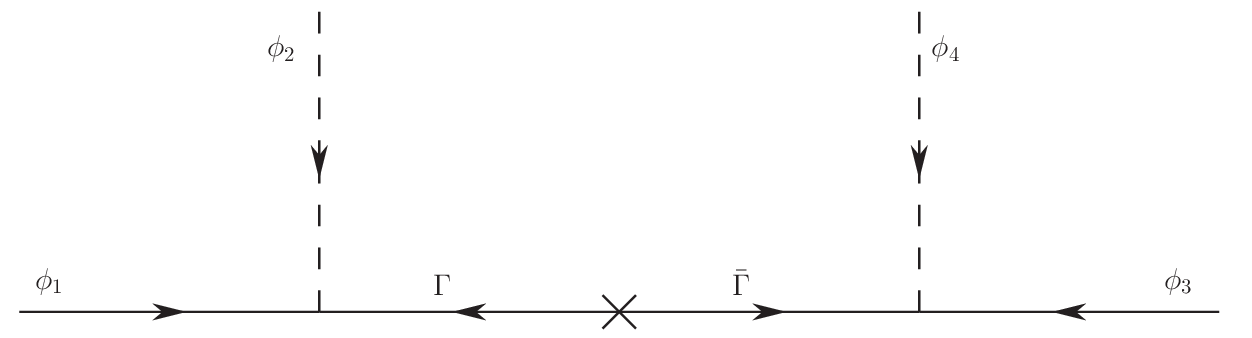}}
  \caption{\small The supergraph for dimension 5 operators in the superpotential
  of down quarks.}
\end{figure}

In SU(5) GUTs, the left and right handed components of up-type quarks reside in the same
representations $T_i$, that is why the Yukawa couplings are of the form
$T_iT_jH_{s}$ ($s=5,45$) which made the Yukawa elements either symmetric or antisymmetric.
However those of down-type quarks live in different representations:
$F_i$ includes the right handed component and $T_i$ has left handed quark doublet.
And vice verse for charged leptons. Then the interactions of both down quarks and charged
leptons are written by the same joint operators of the form $T_iF_jH_{\bar{s}}$.
Accordingly the corresponding Yukawa couplings, $Y_d$ and $Y_e$, are just mutual
transposed relation up to $\mathcal{O}(1)$ CG factors. Therefore the eigenvalues have
to satisfy certain phenomenological GUT relations. For the first two families,
the following double Yukawa ratio\cite{Antusch/2013RUN}
\begin{equation}\label{eq:doubleratio}
\frac{y_{\mu}}{y_s}(\frac{y_e}{y_d})^{-1}=10.7^{+1.8}_{-0.8},
\end{equation}
is a strong restriction on model building. Besides the Yukawa eigenvalues in both
up and down quarks sectors, the quark mixing CKM parameters have to be fulfil as
well. All the requirements imply that only certain CG factors can be realistic in
GUT flavor models. To be specific, in our model the effective superpotential
in down quarks sector and charged leptons sector is written by the operators
\begin{equation}\label{eq:wdownII}
  \mathcal{W}_d=
  \frac{\alpha_d}{\Lambda}[T_1H_{\overline{45}}]_{45}[FH_{24}]_{\overline{45}}Y_{\bf{2}}+
  \frac{\beta_d}{\Lambda}[T_2H_{24}]_{10}[FH_{\bar{5}}]_{\overline{10}}Y_{\bf{2}}+
  \frac{\gamma_d}{\Lambda}[T_3H_{\bar{5}}]_5[F_3H_{24}]_{\bar{5}}+
  g_dT_1F_3H_{\overline{45}},
\end{equation}
where the SU(5) contraction $[XY]_R$ of fields $X$ and $Y$ denotes a tensor in
the representation $R$. Here the adjoint scalar $H_{24}$ is crucial to form the dimension
five operators which result the new Yukawa ratios between leptons and quarks. After the
GUT gauge symmetry is broken when $H_{24}$ develops its VEV along the hypercharge
direction, the Clebsch-Gordan factors emerge in the entries of lepton and quark Yukawa
matrices. These SU(5) tensor contractions are realized by integrating out
heavy messengers. The effective operators in the superpotential are then generated,
see Figure.~\hyperref[fig:TFHH]{2}. For the model to work, we list the messenger fields
and their representations as well as weights in Table~\ref{ta:TFHH}.

\begin{figure}[htb]
  \centering
  \subfigure[]{
  \label{fig:subfig:2H24} %% label for first subfigure
  \includegraphics[width=2.7in]%[bb=0 0 260 175,width=3.2in]
  {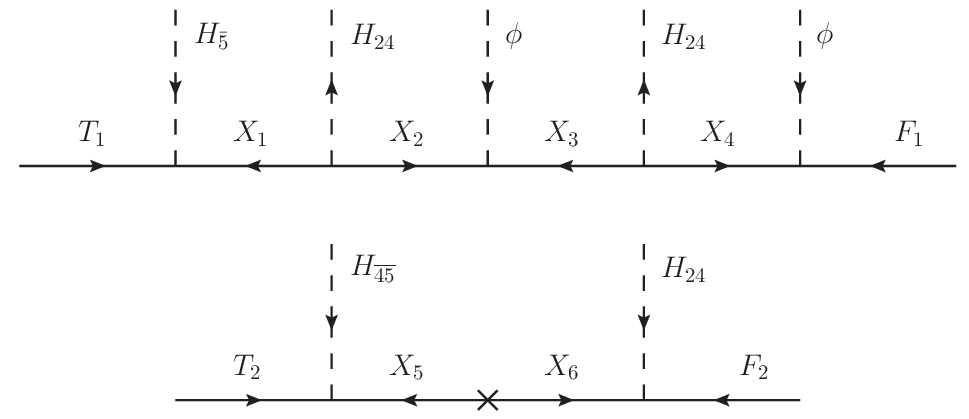}}
  \hskip0.1in
  \subfigure[]{\label{fig:subfig:3H24}
  \includegraphics[width=3.6in]%[bb=0 0 260 175,width=3.2in]
  {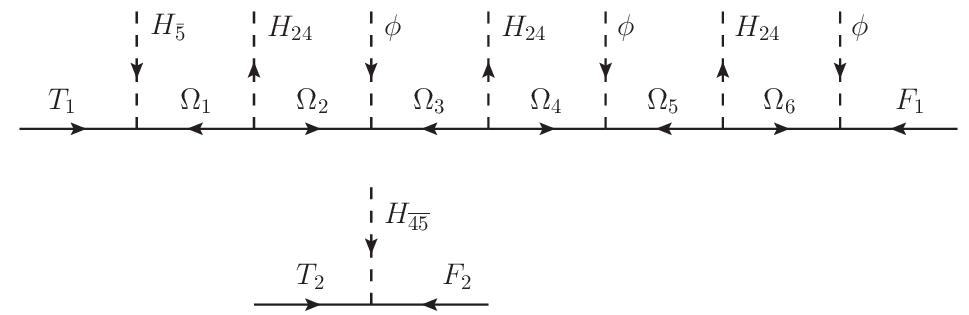}
  }
  \caption{\small The supergraphs for the operators in Eq.~\eqref{eq:caseAII} (left)
  and Eq.~\eqref{eq:caseBII} (right).}\label{fig:MgH24}
\end{figure}

We define the quantity $\epsilon=\langle H_{24}\rangle/\Lambda$, then
the mass matrix of down-type quarks reads
\begin{equation}\label{eq:MdII}
  M_d=\left[
  \begin{array}{ccc}
  \alpha_dv_{\overline{45}}\epsilon Y_2 &
  -\alpha_dv_{\overline{45}}\epsilon Y_1 &
  g_dv_{\overline{45}} \\
  \beta_dv_{\bar{5}}\epsilon Y_2 & \beta_dv_{\bar{5}}\epsilon Y_1 & 0 \\
  0 & 0 & \gamma_dv_{\bar{5}}\epsilon \\
  \end{array}
  \right],
\end{equation}
and that of charged leptons is simply the transposed $M_d$ up to the
CG coefficients of Yukawa couplings
\begin{equation}\label{eq:MeII}
  M_e=\left[\begin{array}{ccc}
  C_1\alpha_dv_{\overline{45}}\epsilon Y_2
  & C_2\beta_dv_{\bar{5}}\epsilon Y_2 & 0 \\
  -C_1\alpha_dv_{\overline{45}}\epsilon Y_1 & C_2\beta_dv_{\bar{5}}\epsilon Y_1 & 0 \\
   C_4g_dv_{\overline{45}} & 0 & C_3\gamma_dv_{\bar{5}}\epsilon \\
  \end{array}
  \right],
\end{equation}
The CG coefficients in the model are $C_1=-1/2$, $C_2=6$, $C_3=-3/2$ and $C_4=-3$.
Note that the fourth CG factor is the famous Georgi-Jarlskog relation
~\cite{Georgi/1979GJ}. The first three ones are the main predictions for the
mass relations between quarks and leptons, which can be realized in conventional models,
such as~\cite{Antusch/2013GUT,Gehrlein/2015GRF}.
We can check that the double ratio in Eq.~\eqref{eq:doubleratio} is satisfied for
the present choice. The set of the CG coefficients, $C_1=-1/2$ and $C_2=6$, is in fact
the only one that can be realized in GUT without flavons. The other two possibilities in
realistic flavor GUT models, e.g., (A) $C_1=4/9, C_2=9/2$~\cite{Bjorkeroth/2015ASU} and
(B) $C_1=-8/27, C_2=-3$~\cite{Zhao/2016GUT}, however, at least one scalar field has to
be added to compensate the loss of mass dimension. In the two cases the mass of
messenger pair $\Gamma$ and $\overline{\Gamma}$ is generated by a heavy scalar field
$\Lambda_{24}$, who lives in the SU(5) adjoint representation. Then the leptonic and
down-type quark-like components of $\Gamma$ and $\overline{\Gamma}$ obtain different
masses split by CG factors. After integrating out the heavy messenger fields, the
CG factors inversely enter in the Yukawa matrix entries of charged leptons
and down-quarks~\cite{Antusch/2014GPN}. We will briefly build
two toy models to elucidate the two cases. For sake of simplicity we assume only one
flavon field $\phi$ (or weighton~\cite{KingSJ/2020FMH})
which is a singlet under a modular symmetry (not necessary
$\Gamma_4$), and the appropriate weight is assigned. And we just show the operators
which result in the diagonal entries of the mass matrix. For case (A) we may write
the superpotential for the first two families as
\begin{equation}\label{eq:caseAII}
  \mathcal{W}^A_d=y^d_{11}T_1F_1H_{\bar{5}}\frac{\phi^2}{H^2_{24}}Y^{(k)}_{\textbf{r},a}
  +\frac{y^d_{22}}{\Lambda}[T_2H_{45}]_{\bar{5}}[F_2H_{24}]_5Y^{(k')}_{\textbf{r}',a'},
\end{equation}
in which $Y^{(k)}_{\textbf{r},a}$ and  $Y^{(k')}_{\textbf{r}',a'}$ denote the components
of modular form multiplets. For case (B) the superpotential is
\begin{equation}\label{eq:caseBII}
  \mathcal{W}^B_d=y^d_{11}T_1F_1H_{\bar{5}}\frac{\phi^3}{H^3_{24}}Y^{(k)}_{\textbf{r},b}
  +y^d_{22}T_2H_{45}F_2Y^{(k')}_{\textbf{r}',b'}.
\end{equation}
The effective operators of the superpotentials are generated after integrate out
the heavy messenger fields. We show in Figure.~\hyperref[fig:subfig:2H24]{3(a)} and
Figure.~\hyperref[fig:subfig:3H24]{3(b)} the supergraphs corresponding to case (A)
and case (B) respectively. The details for building such models are beyond the
scope of this work.

\begin{table}[!htbp]\footnotesize
\renewcommand\arraystretch{1.2}
\centering
\begin{tabular}{|cccccc|}
  \hline
  \hline
  Field & $\phi_1$ & $\phi_2$ & $\phi_3$ & $\phi_4$ & $\Gamma$ \\
  \hline
  1 & $T_1$ & $H_{\overline{45}}$ & $F$ & $H_{24}$ & $\Upsilon_1$  \\
  2 & $T_2$ & $H_{24}$ & $F$ & $H_{\bar{5}}$ & $\Upsilon_2$  \\
  3 & $T_3$ & $H_{\bar{5}}$ & $F$ & $H_{24}$ & $\Upsilon_3$ \\
  \hline
  \hline
\end{tabular}
\hskip0.1cm
\begin{tabular}{|cccc|}
  \hline
  \hline
  Field & $\Upsilon_1$ & $\Upsilon_2$ & $\Upsilon_3$ \\
  \hline
  SU(5) & $\overline{\textbf{45}}$ & $\overline{\textbf{10}}$ & $\bar{\textbf{5}}$  \\
  $\Gamma_4\equiv S_4$ & 1 & $1'$ & $1'$ \\
  $k_I$ & $-1$ & 3 & $-2$ \\
  \hline
  \hline
\end{tabular}
\caption{Left panel: The dimension 5 operators corresponding to
  ~Figure.~\hyperref[fig:TFHH]{2},
  Right panel: The charges of messenger fields appear in the dimension 5 operators}
  \label{ta:TFHH}
\end{table}

\subsection{Neutrino}\label{subsec:neutrino}
In this section we shall give the neutrino interactions which induce light
neutrino masses. We assume neutrinos to be Majorana type, and the light masses
are generated by seesaw mechanism. In most of typical flavor models, including
the flavor GUT models, the light neutrino masses can be produced through type-I
seesaw mechanism which demands at least two superheavy right-handed Majorana
neutrinos to suppress the Yukawa couplings. In the implementation ways of
seesaw mechanisms in SU(5) GUTs, however, there are another two ways to do the
same thing, the first scheme is type-II seesaw
~\cite{Lazarides/1981SGT,Schechter/1980NUM,Mohapatra/1981NMV} by adding an
extra Higgs $\textbf{15}_H$, and the second one is Type-I plus Type-III
seesaw~\cite{Foot/1989SS3} by introducing the fermionic fields in the
\textbf{24} dimensional representation. In our model we assume the second scheme
as the unique origin of neutrino masses and no more extra matter fields are involved.
The matter chiral superfield $A$ lives in SU(5) adjoint representation $\textbf{24}$ and
is also an $S_4$ triplet. According to the symmetry constraints in , the neutrino
Yukawa interactions reads
\begin{equation}\label{eq:nuYukawaII}
\mathcal{W}_{Y_A}=\frac{y_{\nu_1}}{\Lambda}FAH_5Y_{\bf{3}'}
+\frac{y_{\nu_2}}{\Lambda}F_3AH_5Y_{\bf{3}'}.
\end{equation}
Note that the second term would vanish if $(\rho_{F_3},\rho_{A})
=(\textbf{1},\textbf{3})$ or ($\textbf{1}',\textbf{3}'$). If we made the choice,
the third column of Yukawa texture will be vanishing which features a zero determinant
of neutrino mass matrix, no matter what structure of Majorana mass matrix is. We drop
the case at present. The Yukawa matrix is then
\begin{eqnarray}\label{eq:YukawaYII1}
& &  \mathcal{Y}_{\rho_3}=
\frac{y_{\nu_1}}{2}\left[
\begin{array}{ccc}
Y_4 & Y_5 & 0 \\
Y_3 & Y_4 & 0 \\
Y_5 & Y_3 & 0 \\
\end{array}
\right]+\frac{y_{\nu_2}}{2}\left[
\begin{array}{ccc}
0 & 0 & Y_3 \\
0 & 0 & Y_5 \\
0 & 0 & Y_4 \\
\end{array}
\right]\nonumber\\
& & \mathcal{Y}_{\rho_0}=\sqrt{\frac{3}{5}}\mathcal{Y}_{\rho_3}.
\end{eqnarray}

If we introduce SU(5) singlet $\textbf{1}$, e.g., an $N^c$, as right-handed
Majorana neutrinos, the pure mass term of the form $MN^cN^c$ is the only interaction.
However in adjoint SU(5) there is an extra interaction form besides pure mass term.
Since we have $\textbf{24}$ dimensional matter superfield responsible for the
generation of neutrino masses, the new interactions between $\textbf{24}$
and $\textbf{24}_H$ have to be considered.
Now we give all the mass terms of $\textbf{24}$ as
\begin{equation}\label{eq:nuMajoranaII}
\mathcal{W}_A=MAAY_{\bf{2}}+M'AAY_{\bf{3}'}
+\lambda_1AAH_{24}Y^{(4)}_{\bf{2}}+\lambda_2AAH_{24}Y^{(4)}_{\bf{3'}}
+\lambda_3AAH_{24}Y^{(4)}_{\bf{3}}.
\end{equation}
The second  and the fifth terms vanish because the tensor productions
of $\textbf{3}'\otimes\textbf{3}'$ have to be antisymmetric
(see~\hyperref[A]{Appendix}) to form an invariant with modular forms.
Appling the decomposition to eq.~\eqref{eq:nuMajoranaII},
the fermionic singlet $\rho_0$ and triplet $\rho_3$
have the mass matrices of the form
\begin{eqnarray}\label{eq:MrhoII}
& &  M_{\rho_3}= \nonumber\\
& & \frac{M}{4}
\left[
\begin{array}{ccc}
0 & Y_1 & Y_2 \\
Y_1 & Y_2 & 0 \\
Y_2 & 0 & Y_1 \\
\end{array}
\right]
-\frac{3v'_{24}}{4\sqrt{30}}\Big\{\lambda_1
\left[
\begin{array}{ccc}
0 & -Y^2_2 & Y^2_1 \\
-Y^2_2 & Y^2_1 & 0 \\
Y^2_1 & 0 & -Y^2_2 \\
\end{array}
\right]
+\lambda_2
\left[
\begin{array}{ccc}
2Y^{(4)}_{\textbf{3}',1} & -Y^{(4)}_{\textbf{3}',3} & -Y^{(4)}_{\textbf{3}',2} \\
-Y^{(4)}_{\textbf{3}',3} & 2Y^{(4)}_{\textbf{3}',2} & -Y^{(4)}_{\textbf{3}',1} \\
-Y^{(4)}_{\textbf{3}',2} & -Y^{(4)}_{\textbf{3}',1} & 2Y^{(4)}_{\textbf{3}',3} \\
\end{array}
\right]\Big\}, \nonumber\\
& &  M_{\rho_0}= \\
& &  \frac{M}{4}
\left[
\begin{array}{ccc}
0 & Y_1 & Y_2 \\
Y_1 & Y_2 & 0 \\
Y_2 & 0 & Y_1 \\
\end{array}
\right]
-\frac{v'_{24}}{4\sqrt{30}}
\Big\{\lambda_1
\left[
\begin{array}{ccc}
0 & -Y^2_2 & Y^2_1 \\
-Y^2_2 & Y^2_1 & 0 \\
Y^2_1 & 0 & -Y^2_2 \\
\end{array}
\right]
+\lambda_2
\left[
\begin{array}{ccc}
2Y^{(4)}_{\textbf{3}',1} & -Y^{(4)}_{\textbf{3}',3} & -Y^{(4)}_{\textbf{3}',2} \\
-Y^{(4)}_{\textbf{3}',3} & 2Y^{(4)}_{\textbf{3}',2} & -Y^{(4)}_{\textbf{3}',1} \\
-Y^{(4)}_{\textbf{3}',2} & -Y^{(4)}_{\textbf{3}',1} & 2Y^{(4)}_{\textbf{3}',3} \\
\end{array}
\right]\Big\},\nonumber
\end{eqnarray}
where the abbreviation $Y^{(4)}_{\textbf{3}',i}$($i=1,2,3$) denote the components of
weight 4 modular form $Y^{(4)}_{\textbf{3}'}$ for the expression more compact,
e.g., $Y^{(4)}_{\textbf{3}',1}=Y_1Y_4+Y_2Y_5$ and else, see Eq.~\eqref{eq:Yw4}.

The neutrino masses are generated by Type-I and Type-III seesaw, which are realized
by integrating out $\rho_0$ and $\rho_3$, respectively. We write
the effective light neutrino mass matrix as the sum of the contributions from
Type-I and Type-III seesaw
\begin{equation}\label{eq:MnuSSII}
M^{SS}_{\nu}=
-(\mathcal{Y}^T_{\rho_0}M^{-1}_{\rho_0}\mathcal{Y}_{\rho_0}
+\mathcal{Y}^T_{\rho_3}M^{-1}_{\rho_3}\mathcal{Y}_{\rho_3})v^2_u,
\end{equation}
where $v_u=v\sin\beta$ with $v=\sqrt{v^2_u+v^2_d}=$174GeV and $\tan\beta=v_u/v_d$,
as usual defined in MSSM.

\section{Phenomenology}\label{sec:Phenomenology}
The superpotential in the matter sector is simply given by
\begin{equation}\label{eq:Wmatter}
  \mathcal{W}_{\textrm{matt}}=%+\mathcal{W}_{\nu}
  \mathcal{W}_u+\mathcal{W}_d+\mathcal{W}_{Y_A}+\mathcal{W}_A,
\end{equation}
in which each term is given by Eqs.~\eqref{eq:wupII},~\eqref{eq:wdownII},
~\eqref{eq:nuYukawaII} and~\eqref{eq:nuMajoranaII}, respectively. As the GUT and
modular symmetry breaking we write the Yukawa matrices of fermions
in the following convention
\begin{eqnarray}\label{eq:WYukawa}
& & W_Y=(Y_u)_{ij}q_i\ast H_uu^c_j+(Y_d)_{ij}q_i\ast H_dd^c_j
+(Y_e)_{ij}\ell_i\ast H_de^c_j\nonumber\\
& & +(\mathcal{Y}_{\rho_3})_{ij}\ell^T_ii\sigma_2(\rho_3)_jH_u
+(\mathcal{Y}_{\rho_0})_{ij}\ell^T_ii\sigma_2(\rho_0)_jH_u
+(M_{\rho_3})_{ij}\textrm{Tr}((\rho_3)_i(\rho_3)_j)+(M_{\rho_0})_{ij}(\rho_0)_i(\rho_0)_j,
\end{eqnarray}
where $H_u$ and $H_d$ are the two Higgs doublets in MSSM. The neutrino Yukawa matrices
and mass matrices have been given by Eqs.~\eqref{eq:YukawaYII1} and~\eqref{eq:MrhoII}
in Sec.~\ref{subsec:neutrino}. The Yukawa matrices of charged fermion\textbf{s} are
then read as
\begin{equation}\label{eq:Yu}
  Y_u=\left[
  \begin{array}{ccc}
  \alpha_u & 0 & 0 \\
  0 & \beta_uY_1Y_2 & g_u \\
  0 & -g_u & \gamma_u \\
  \end{array}
  \right]\equiv\left[
  \begin{array}{ccc}
  u_{11} & 0 & 0 \\
  0 & u_{22} & u_{23} \\
  0 & -u_{23} & u_{33} \\
  \end{array}
  \right],
\end{equation}
for up quarks, and
\begin{equation}\label{eq:Yd}
  Y_d=\left[
  \begin{array}{ccc}
  \alpha_dY_2 & -\alpha_dY_1 & g_d \\
  \beta_dY_2 & \beta_dY_1 & 0 \\
  0 & 0 & \gamma_d \\
 \end{array}\right]\equiv
 \left[
   \begin{array}{ccc}
     d_{11} & -d_{11}Y_1/Y_2 & d_{13} \\
     d_{22}Y_2/Y_1 & d_{22} & 0 \\
     0 & 0 & d_{33} \\
   \end{array}
 \right],
\end{equation}
for down quarks. Observing the Yukawa matrices $Y_u$ and $Y_d$ are very sparse
with some texture zeros, one can conclude the CKM Cabibbo angle $\theta^q_{12}$ is
totally generated by the mixing in the down sector, and of course the same for
the $\theta^q_{13}$ as well as CP phase $\delta^q$. The mixing in up sector completely
determines the angle $\theta^q_{23}$.

In SU(5) GUT the Yukawa matrix of charged leptons is the transpose
of that of down quarks, i.e., $Y_e\sim Y^T_d$, up to the order one CG coefficients,
\begin{equation}\label{eq:Ye}
  Y_e=\left[
        \begin{array}{ccc}
          -\frac{1}{2}d_{11} & 6d_{22}Y_2/Y_1 & 0 \\
          -\frac{1}{2}d_{11}Y_1/Y_2 & 6d_{22} & 0 \\
          -3d_{13} & 0 & -\frac{3}{2}d_{33} \\
        \end{array}
      \right].
\end{equation}
It is obvious that the down quarks and charged leptons follow the following
Yukawa ratios
\begin{equation}\label{eq:ratios}
  \frac{y_{\tau}}{y_b}\approx\frac{3}{2},\quad
  \frac{y_\mu}{y_s}\approx6,\quad
  \frac{y_e}{y_d}\approx\frac{1}{2},
\end{equation}
in which $y_{\ell}$ ($\ell=e,\mu,\tau$) and $y_{d_i}$ ($i=d,s,b$) are the eigenvalues of
$Y_e$ and $Y_d$, respectively. Thus the CG factors $C_1=1/2$ and $C_2=6$ made the
double Yukawa ratio $y_dy_{\mu}/y_ey_s=12$ which is in good agreement with the data 
in Eq.~\eqref{eq:doubleratio}.

\subsection{SUSY Threshold corrections}\label{subsec:threshold}
The Yukawa couplings and mixing observables are defined at superhigh GUT scale in our
model, therefore the values from low energy experiments must run up to the ones at
GUT scale. Moreover the SUSY radiative threshold corrections
are the requisite factor for matching the MSSM at the SUSY scale $M_{SUSY}$
to the SM~\cite{Hall/1994THS,Carena/1994YTH,Hempfling/1994UTH,Blazek/1995FTH}.
The running of MSSM Yukawa parameters from $M_Z$ to $M_{GUT}$ has been
analysed in~\cite{Antusch/2013RUN}, where the $\tan\beta$ enhanced 1-loop SUSY
threshold effects are discussed in detail. The matching relations between the
eigenvalues of the MSSM and the SM Yukawa coupling matrices are parameterized as
\begin{eqnarray}\label{eq:matching}
& &  y^{\textrm{MSSM}}_{u,c,t}=\frac{y^{\textrm{SM}}_{u,c,t}}{\sin\bar{\beta}},\\
& &  y^{\textrm{MSSM}}_{d,s}
=\frac{y^{\textrm{SM}}_{d,s}}{(1+\bar{\eta}_q)\cos\bar{\beta}},\qquad
 y^{\textrm{MSSM}}_b
=\frac{y^{\textrm{SM}}_b}{(1+\bar{\eta}_b)\cos\bar{\beta}}, \\
& &  y^{\textrm{MSSM}}_{e,\mu}
=\frac{y^{\textrm{SM}}_{e,\mu}}{(1+\bar{\eta}_\ell)\cos\bar{\beta}},\qquad
 y^{\textrm{MSSM}}_{\tau}
=\frac{y^{\textrm{SM}}_{\tau}}{\cos\bar{\beta}},
\end{eqnarray}
and the quark CKM parameters are also corrected by
\begin{equation}\label{eq:CKMparameters}
\theta^{q,\textrm{MSSM}}_{i3}=\frac{1+\bar{\eta}_b}{1+\bar{\eta}_q}\theta^{q,\textrm{SM}}_{i3},\quad
\theta^{q,\textrm{MSSM}}_{12}=\theta^{q,\textrm{SM}}_{12},\quad
\delta^{q,\textrm{MSSM}}=\delta^{q,\textrm{SM}}.
\end{equation}
One can notice, to a good approximation, the threshold corrections have no impact on
$\theta^q_{12}$ and $\delta^q$ and the running of Yukawa couplings $Y^{\textrm{MSSM}}_f$
depends only on $\bar{\eta}_b$ and $\tan\bar{\beta}$. Especially in the limit that
the threshold corrections to charged leptons are neglected, i.e., $\bar{\eta}_\ell=0$,
then $\tan\bar{\beta}$ reduces to the usual $\tan\beta$. We will adopt the scenario
in the model. Nevertheless the $\bar{\eta}_q$ can not be dropped. The reason is that
the ratio $y_{\mu}/y_s$ at GUT scale is approximately 4.5 without SUSY threshold
corrections, but the large CG factor 6 in \eqref{eq:Ye} needs a compensation from
$\bar{\eta}_q$ which is approximately +0.33~\cite{Antusch/2018CSD}.
The two CG factors $C_2=6$ and $C_3=-3/2$
appeared in the Yukawa matrix $Y_e$ in Eq.~\eqref{eq:Ye} require a relative large
$\tan\beta$ to generate substantial threshold corrections. In fact for the large
Yukawa coupling ratios in \eqref{eq:ratios}, both large $\tan\beta$ and large
threshold corrections are required~(cf.\cite{Antusch/2009NGUT}). Accordingly
we set $\tan\beta=35$, $\bar{\eta}_b=0.13125$ and $\bar{\eta}_{q}=0.3$. We notice
that the threshold parameters can be free, but the fixed values are enough to
reproduce correct observables.

\subsection{Numerical Analysis}\label{subsec:Numerical}
The Yukawa matrices for neutrinos, up- and down- quarks and leptons are presented
in Eqs.~\eqref{eq:MnuSSII}, \eqref{eq:Yu}, \eqref{eq:Yd} and \eqref{eq:Ye}, respectively.
The only common parameter $\tau$ among them has been bounded in the upper half complex
plane. Modular symmetry itself, however, is enable to give rise to the hierarchical
fermion masses, and mixing parameters. The free parameters appear in the mass matrices are
\begin{equation}\label{eq:freeparameters}
  P_i=\{\tau,\alpha_u,\beta_u,\gamma_u,g_u,\alpha_d,\beta_d,\gamma_d,g_d,
  y_{\nu_1},y_{\nu_2},M,\lambda_1,\lambda_2,\},
\end{equation}
and the physical observables $Q^{\textrm{obs}}$ in the GUT model include
\begin{eqnarray}\label{eq:observable}
& &  Q_q=\{y_u,y_c,y_t,y_d,y_s,y_b,
    \theta^q_{12},\theta^q_{13},\theta^q_{23},\delta^q\},\nonumber\\
& &  Q_{\ell}=\{y_e,y_{\mu},y_{\tau},
   \Delta \mathcal{M}^2_{21}, \Delta \mathcal{M}^2_{31},
   \theta^{\ell}_{12},\theta^{\ell}_{13},\theta^{\ell}_{23},\delta^{\ell}\}.
\end{eqnarray}
We shall construct a global $\chi^2$ function for all the observable quantities when
fitting the coupling matrices in~Eqs.\eqref{eq:Yu}, \eqref{eq:Yd} and \eqref{eq:Ye}
for charged fermions and Eqs.~\eqref{eq:YukawaYII1} and \eqref{eq:MrhoII} for neutrinos.
The $\chi^2$ function to be minimized is defined as
\begin{equation}\label{eq:chi2}
  \chi^2=\sum_i\Big(\frac{Q_i(P_j)-Q^{\textrm{obs}}_i}{\sigma_i}\Big)^2,
\end{equation}
where $Q_i(P_j)$ denote the model predicted values for observables and
$Q^{\textrm{obs}}=Q_q+Q_{\ell}$ are the central values with $\sigma_i$
the $1\sigma$ errors.

Before performing the fit, we would like to elucidate the data used in the minimization.
As stressed in Sec.~\ref{subsec:threshold}, the model is defined at the high
energy scale, the observable quantities should be set at the GUT scale.
The quantities at GUT energy scale can be achieved from the
low energy scale where the experimental values are determined by the renormalization
group equations (RGEs). Here in our work we adopt the values of Yukawa couplings
at the GUT scale $M_X=2\times 10^{16}$GeV, assuming minimal SUSY breaking
$M_{SUSY}=$1TeV with a large $\tan\beta=35$~\cite{Antusch/2013RUN}. The Yukawa values
$y_f$ give rise to the fermion masses as $m_f=y_fv_H$ with $v_H$=174GeV. Meanwhile
the CKM mixing angles and CP phase are also taken as the values at the GUT scale.
All the Yukawa values and CKM observables as well as their $1\sigma$ errors are
listed in the left panel of Table~\ref{ta:value}.
For the neutrino sector, the lepton mixing angles, CP phase and neutrino mass squared
differences we adopted are taken from NuFit 5.0~\cite{Esteban/2020GNF}. We show
their center values and $1\sigma$ errors in the right panel of
Table~\ref{ta:value}. The above input data are used for the estimation
of our $\chi^2$.

Since the quark sector and neutrino sector have the modulus $\tau$ as the common
parameter which determines the modular forms, the remaining free parameters are
just Yukawa coupling coefficients. It is equivalent and more convenient to fit the
entries of Yukawa matrices rather than Yukawa coefficients themselves. Therefore
we just fit the entries of quark Yukawa matrices, $u_{ij}$s in~Eq.~\eqref{eq:Yu} and
$d_{ij}$s in~Eq.~\eqref{eq:Yd}. In neutrino sector, we define the coupling
parameter ratios
\begin{equation}\label{eq:nuYukawaRatios}
  r_{21}=\frac{y_{\nu_2}}{y_{\nu_1}},\qquad
  r_1=\frac{\lambda_1v_{24}}{M},\qquad
  r_2=\frac{\lambda_2v_{24}}{M},
\end{equation}
and an overall mass scale $y^2_{\nu_1}v^2_u/M$ are the parameters to be fitted.
So, instead of fitting the primitive free parameters showed
in Eq.~\eqref{eq:freeparameters}, we take the following equivalent parameter set
\begin{equation}\label{eq:freeparameters2}
  \overline{P}_i=\{\tau,u_{11},u_{22},u_{33},u_{23},d_{11},d_{22},d_{33},d_{13},
  r_{21},r_1,r_2,y^2_{\nu_1}v^2_u/M\}.
\end{equation}

In Table~\ref{ta:input} we show the model best fit input parameters
in quark Yukawa and neutrino mass matrices which minimizes the $\chi^2_q$
and $\chi^2_{\ell}$, respectively. We present our fit results for all the
Yukawas and mass matrices in Table~\ref{ta:fit}. %and Table~\hyperref[ta:fit2]{6}.
In the left panel of Table~\ref{ta:fit} we give the resulting best fit values
and pulls to the ten quark observables: three quark CKM mixing angles
$\theta^q_{ij}$ ($ij=12,13,23$) and one CP violating phase $\delta^q$, six Yukawas
$y_q$ ($q=u,c,t,d,s,b$). Also the minimum $\chi^2_{\textrm{min},q}\sim0.45$ is
given at the last row. We also list in Table~\ref{ta:fit} (right panel)
the best fit values and pulls to six neutrino observables and three charged
lepton Yukawas. The minimum $\chi^2_{\textrm{min},\ell}$ is just $\mathcal{O}(1)$.
The best fit point has the total
$\chi^2_{\textrm{min}}=\chi^2_{\textrm{min},q}+\chi^2_{\textrm{min},\ell}\simeq1.6$.
One can see that the model favours normal ordering neutrino masses, and we found
the minimum $\chi^2_{\textrm{min},\ell}=8.478$ for inverted ordering. Besides the
values of absolute neutrino masses $m_i$, the Majorana phases $\varphi_{21}$ and
$\varphi_{31}$ are pure theoretical predictions. The mass sum,
the $\beta$-decay effective mass $m_{\beta}$ as well as the neutrinoless
double beta ($0\nu\beta\beta$) decay
amplitude parameter $m_{ee}$ are also given as predictions in the table. Specifically
the bounds on the above mass related quantities are given by
\begin{eqnarray}\label{eq:masssum}
& &  \sum_im_i\leq120 \textrm{meV}, \nonumber\\
& &  m_{\beta}=\Big[\sum_im^2_i|U_{ei}|^2\Big]^{1/2}<(61\sim165)\textrm{meV},\\
& &  m_{ee}=\Big|\sum_iU^2_{ei}m_i\Big|<1.1\textrm{eV}(90\%\textrm{C.L.}),\nonumber
\end{eqnarray}
which are taken from~PLANCK\cite{PlanckCollaboration/2020CVI},
KamLAND-ZEN~\cite{Gando/2016KLZ} and KATRIN~\cite{Aker/2019KTR}, respectively.
The leptonic mixing matrix elements $U_{ei}$ are taken from the standard
parametrization~\cite{Tanabashi/2018PDG}
\begin{equation}\label{eq:UMNS}
  U=\left(
  \begin{array}{ccc}
  c_{12}c_{13} & s_{12}c_{13} & s_{13}e^{-i\delta} \\
  -s_{12}c_{23}-c_{12}s_{13}s_{23}e^{i\delta} &
  c_{12}c_{23}-s_{12}s_{13}s_{23}e^{i\delta} & c_{13}s_{23} \\
  s_{12}s_{23}-c_{12}s_{13}c_{23}e^{i\delta} &
  -c_{12}c_{23}-s_{12}s_{13}c_{23}e^{i\delta} & c_{13}c_{23} \\
  \end{array}
  \right)
  \left(
    \begin{array}{ccc}
      1 & 0 & 0 \\
      0 & e^{\varphi_{21}/2} & 0 \\
      0 & 0 & e^{\varphi_{31}/2} \\
    \end{array}
  \right),
\end{equation}
where $s_{ij}=\sin\theta_{ij}$, $c_{ij}=\cos\theta_{ij}$, $\delta$ is lepton Dirac CP
violating phase and $\varphi_{21}$, $\varphi_{31}$ are two Majorana CP phases.
Comparing the model predictions in Table~\ref{ta:fit} to these bounds,
we see that the predicted values are well below the corresponding upper bounds.
For the sum of neutrino masses, our result is still below the tightest and
most rebust upper limit $M_{\nu}<0.118$eV~\cite{Vagnozzi/2017NMS}.

There are 17 real input parameters in Table~\ref{ta:input}, to fit 19 measured
data points in Table~\ref{ta:value}. Hence the number of the degree of
freedom (d.o.f) is naively 2, which is the difference of the number of observables
and inputs. This translates to a reduced $\chi^2$, i.e.,
$\chi^2_{\textrm{red}}=\chi^2_{\textrm{min}}/\textrm{d.o.f}=0.808$.
We take this value as a good fit. The fit has been performed by using the
Mathematica package Mixing Parameter Tools (MPT)~\cite{Antusch/2005MPT}.

We can see that in quark and lepton sectors the only common input
parameter is modulus $\tau$, which is close to the boundary (right cusp) of the
fundamental region. The Yukawas are in general complex, however we can always absorb
most of the phases by redefinition of the fields. Specifically for the quark Yukawa
inputs, the only complex parameter is $d_{13}$ which mainly controls the magnitude of
the CKM matrix element $V_{ub}$ and part of the CP phase $\delta^q$. Meanwhile the other
inputs, all $u_{ij}$s and the rest of $d_{ij}$s ($ij\neq13$) are real. In neutrino
sector we also have reduced the total number of real parameters to be six, in which
only $r_1$ and $r_2$ are complex. The Yukawa ratio $r_{21}$ and the overall mass scale
$y^2_{\nu_1}v^2_u/M$ are all real parameters.

\begin{table}[!htbp]%\footnotesize
\renewcommand\arraystretch{1.5}
  \centering
\begin{tabular}{|c|cc|}%\label{ta:valueQL}
  \hline
  %Observable & $\tan\beta=35$ & \\
  Observable  & $\mu_i$ & $\sigma_i$\\
  \hline
  $\theta^q_{12}/^\circ$  & 13.027  & 0.041 \\
  $\theta^q_{13}/^\circ$  & 0.166   &   0.006 \\
  $\theta^q_{23}/^\circ$  & 1.924  & 0.031 \\
  $\delta^q/^\circ$  & 69.213  &  3.094 \\
  \hline
  $y_u$  & $2.889\times10^{-6}$  & $8.954\times10^{-7}$  \\
  $y_c$  & $1.413\times10^{-3}$  & $4.946\times10^{-5}$ \\
  $y_t$  & 0.529 &  0.016  \\
  \hline
  $y_d$  & $1.3879\times10^{-4}$  & $1.5267\times10^{-5}$  \\
  $y_s$  & $2.7467\times10^{-3}$  & $1.4832\times10^{-4}$  \\
  $y_b$  & 0.1867  & $2.766\times10^{-3}$  \\
  \hline
  $y_e$ & $7.3834\times10^{-5}$  & $4.430\times10^{-7}$  \\
  $y_{\mu}$ & $1.5590\times10^{-2}$ & $9.354\times10^{-5}$  \\
  $y_{\tau}$ & 0.2819 & $1.751\times10^{-3}$  \\
  \hline
\end{tabular}
\hskip0.3cm
\begin{tabular}{|c|c|c|}%\label{ta:valueNU}
  \hline
  Observable & $\textrm{NO}$ & $\textrm{IO}$ \\
  \hline
  $\theta_{12}/^\circ$  & $33.44^{+0.78}_{-0.75}$ & $33.45^{+0.78}_{-0.75}$\\
  $\theta_{13}/^\circ$  & $8.57^{+0.13}_{-0.12}$  &   $8.61\pm0.12$ \\
  $\theta_{23}/^\circ$  & $49.0^{+1.0}_{-1.4}$ & $49.3^{+1.0}_{-1.2}$  \\
  $\delta/^\circ$  & $195^{+51}_{-25}$   & $286^{+27}_{-32}$  \\
  \hline
 $\Delta\mathcal{M}^2_{21}/10^{-5}$ & $7.42^{+0.21}_{-0.20}$  &$7.42^{+0.21}_{-0.20}$ \\
  $\Delta\mathcal{M}^2_{31}/10^{-3}$  & $2.514^{+0.028}_{-0.027}$  &
  $-2.497^{+0.028}_{-0.028}$ \\
  \hline
\end{tabular}
\caption{Left panel: The Observables of charged fermions at the GUT scale
for $\tan\beta=35$ are taken from~\cite{Antusch/2013RUN}. The SUSY breaking scale are
set at $M_{SUSY}=$1TeV and the threshold correction parameters $\bar{\eta}_b=0.13125$
and $\bar{\eta}_q=0.3$.
Right panel: The values of neutrino Observables are taken from NuFit5.0
~\cite{Esteban/2020GNF} without the atmospheric data from SuperKamiokande.
NO (IO) denotes the Normal (Inverted) Ordering of neutrino masses.}\label{ta:value}
\end{table}

\begin{table}[!htbp]%\footnotesize
\renewcommand\arraystretch{1.5}
\centering
\begin{minipage}{\textwidth}
\begin{tabular}{|c|c|}
  \hline
  Quark Input & Value\\
  \hline
%  $\tau$ & $0.4768+0.9288i$ \\
%  \hline
  $u_{11}$  & $2.887\times10^{-6}$\\
  $u_{22}$  & $8.476\times10^{-4}$  \\
  $u_{33}$  & $0.5304$ \\
  $u_{23}$  & $-1.775\times10^{-2}$\\
  \hline
  $d_{11}$  & $2.667\times10^{-6}$\\
  $d_{22}$  & $1.020\times10^{-4}$ \\
  $d_{33}$  & $6.032\times10^{-3}$ \\
  $d_{13}$  & $2.031\times10^{-5}\textrm{e}^{-0.657i\pi}$ \\
  \hline
\end{tabular}
\hskip0.3cm
\begin{minipage}{0.4\textwidth}
\begin{tabular}{|c|c|}
  \hline
  Neutrino Input & Value \\
  \hline
  $r_1=\lambda_1v_{24}/M$  & $-0.06587+2.09536i$ \\
  $r_2=\lambda_2v_{24}/M$  & $0.36449+1.72602i$ \\
  $r_{21}=y_{\nu_2}/y_{\nu_1}$  & 2.26385 \\
  \hline
  $y_{\nu_1}^2v^2_u/M(\textrm{eV})$ & 0.00362\\
  \hline
\end{tabular}
%\hskip0.1cm
\vskip0.2cm
\begin{tabular}{|c|c|}
  \hline
  Common Input & Value \\
  \hline
  $\texttt{Re}\tau$ & $0.4768$ \\
  $\texttt{Im}\tau$ & $0.9288$ \\
  \hline
\end{tabular}
\end{minipage}%}
\end{minipage}
\caption{Left panel: The best fit input parameters in quark sector. The Yukawa entries
  $u_{ij}$ and $d_{ij}$ include the original Yukawa coupling coefficients
  and the modular forms.
  Upper Right panel: The best fit input parameters in neutrino sector.
  We list the ratios of neutrino couplings in eqs.~\eqref{eq:nuYukawaII}
  and~\eqref{eq:nuMajoranaII}.
  Lower Right panel: The only common input parameter to both sectors
  is the modulus $\tau$.}\label{ta:input}
\end{table}

\begin{table}[!htbp]%\footnotesize
\renewcommand\arraystretch{1.5}
  \centering
%\subtable{
  \begin{tabular}{|c|ccc|}
  \hline
  Quark output & $\textrm{Value}$ & & $\textrm{pull}$\\
  \hline
  $\theta^q_{12}/^\circ$  & 13.0233 & & $-0.0839$\\
  $\theta^q_{13}/^\circ$  & 0.16639 & &   0.0002 \\
  $\theta^q_{23}/^\circ$  & 1.9238 & & 0 \\
  $\delta^q/^\circ$  & 69.0969  & & 0.0376  \\
  \hline
  $y_u$  & $2.889\times10^{-6}$ & & 0.0  \\
  $y_c$  & $1.413\times10^{-3}$ & & 0.0 \\
  $y_t$  & 0.529 & & 0.0  \\
  \hline
  $y_d$  & $1.4378\times10^{-4}$ & & 0.327  \\
  $y_s$  & $2.6685\times10^{-3}$ & & $-0.527$  \\
  $y_b$  & 0.1873 & & 0.2312  \\
  \hline
  $\chi^2_{\textrm{min},q}$  & & & 0.4467  \\
  \hline
\end{tabular}
%}
\hskip0.2cm
%\subtable{
\begin{tabular}{|c|ccc|}
  \hline
  Lepton output & $\textrm{Value}$ & & $\textrm{pull}$\\
  \hline
  $\theta_{12}/^\circ$  & 33.4916 & & 0.0675\\
  $\theta_{13}/^\circ$  & 8.5344  & &   $-0.285$ \\
  $\theta_{23}/^\circ$  & 49.7665 & & 0.613  \\
  $\delta/^\circ$  & 208.2733 & & 0.349 \\
  %\hline
  $\Delta\mathcal{M}^2_{21}/10^{-5}$ & 7.524 & & 0.4966  \\
  $\Delta\mathcal{M}^2_{31}/10^{-3}$  & 2.514 & & 0 \\
%  $r$ & 0.0299 & &
%  \textcolor[rgb]{1.00,0.00,0.00}{0.348} \\
  %\hline
  $y_e$  & $7.377\times10^{-5}$ & & $-0.1432$\\
  $y_{\mu}$  & $1.560\times10^{-2}$ & & 0.1433\\
  $y_{\tau}$  & 0.281 & & $-0.5478$\\
  \hline
  $\chi^2_{\textrm{min},\ell}$  & & & 1.1694\\%/\textcolor[rgb]{1.00,0.00,0.00}{1.04412}\\
  \hline
  Predictions & Value & &\\
  \hline
  Mass Ordering & NO & &\\
  $m_1/\textrm{meV}$ & 11.728 & &\\
  $m_2/\textrm{meV}$ & 14.585 & &\\
  $m_3/\textrm{meV}$ & 51.493 & &\\
  $\sum_im_i/\textrm{meV}$ & 77.809 & &\\
  $m_{\beta}/\textrm{meV}$ & 14.674 & &\\
  $m_{ee}/\textrm{meV}$ & 11.040 & &\\
  $\varphi_{21}/\pi$ & 0.105 & &\\
  $\varphi_{31}/\pi$ & 1.323 & &\\
  \hline
\end{tabular}
%}
\caption{Left panel: The quark fit output results. All the quark sector is fitted
  to the $1\sigma$ interval.
Right panel: The output results of lepton sector. The absolute values of three
light neutrino masses and the ordering as well as the two Majorana phases are
pure model predictions. We have found a minimum $\chi^2_{\textrm{min},\ell}=8.478$
for inverted ordering.
}\label{ta:fit}
\end{table}

\section{Summary}\label{sec:Summary}
In the study we explored an supersymmetric adjoint SU(5) GUT flavor model based on modular
$\Gamma_4 \simeq S_4$ symmetry. We have shown the model can produce
correct masses and mixing parameters of both quarks and leptons simultaneously.
No flavons are introduced to the model, only the  complex modulus $\tau$ is responsible
for the breaking of modular symmetry. By assigning suitable representations and weights
to chiral superfields, only finite coupling terms are presented in the effective
operators in quark sector and lepton sector. We obtained very sparse Yukawa matrices of
up- and down-quarks (and of course charged leptons) with some texture zeros. Also in
neutrino sector we have only two Yukawa coupling terms, and three terms in Majorana
mass terms. The effective light neutrino masses are generated through Type-I plus
Type-III seesaw mechanism. The modulus $\tau$ is the only common field appeared in both
quark and lepton sector as spurion.

For simplicity we have used modular forms with lower weight in the model,
since higher weight would bring more free parameters. The assignments of representations
under $S_4$ and weights for the chiral superfields are also highly constrained such
that the Yukawa coupling terms are uniquely fixed by the modular forms.
The \textbf{10} dimensional matter fields are all $S_4$ singlets,
while $\bar{\textbf{5}}$s are divided into a $S_4$ doublet for the first two generations
and a singlet for third one. Meanwhile the adjoint superfields \textbf{24} are collected
in the triplet of $S_4$. All the scalars of course transform as singlets of $S_4$.
We also assigned distinct weights for the superfields such that the number of free Yukawa
parameters is as little as possible.

With the delicate representation assignments of field content, the superpotentials
of the model is relatively simple. Unlike the models in
~\cite{Anda/2020GAF,Kobayashi/2020SRG} which have more parameters than observables,
our model has less free parameters and thus more predictive than the above two works.
The resulting operators in each sector have limit coupling parameters.
The model predicts that down quarks and charged leptons have a rigid Yukawa
coupling ratios generated by CG factors which arise from the SU(5) contractions of the
effective 5D operators. The double Yukawa ratio $y_{\mu}y_d/y_ey_s$ equal to 12
for the first two families of charged leptons and down quarks. The model has
only 17 real parameters in total, and 19 observables to be fit, thus the degree of
freedom (d.o.f) is 2. We obtained the reduced chi-square
$\chi^2/\textrm{d.o.f}\simeq 0.81$ for quark and lepton sectors combined,
which is a good fit. The model favours an normal
mass ordering over inverted ordering for light neutrinos, with a $\Delta\chi^2\sim7$.
Moreover we obtained the absolute values of three light neutrino masses,
the effective masses of $\beta$-decay and $0\nu\beta\beta$ decay as well as the Majorana
CP phases as model predictions.

At last we give a outlook for the model. Since the masses of the fields live
in the adjoint representation are splited by the adjoint scalar $H_{24}$,
the lightest one is $\rho_3$. It is crucial to realize the baryogenesis via leptogenesis
~\cite{Blanchet/2008BVL,Kannike/2011NLG} in the context.
In the case the net asymmetry of $B-L$ can be generated in the out of equilibrium decays
of $\rho_3$ and $\rho_0$ as well as their superpartners in the adjoint representation.
So it is worth studying further the phenomenology according to the model.

\vskip 0.3cm
\section*{Acknowledgements}
We would like to thank Dr.~Shu-Jun~Rong for useful discussions.
This work is supported by the National Natural Science Foundation of China (NSFC)
under Grant No.11875327, the Fundamental Research Funds for the Central
Universities, China, and the Sun Yat-Sen University Science Foundation.

%%%%%%%%%%%%%%%%%%%%%%%%%%%%%%%%%%%%%%%%%%%%%%%%%%%%%%
\appendix
\begin{appendix}
\renewcommand{\theequation}{{A}\thesection\arabic{equation}}
\setcounter{equation}{0}
\section*{APPENDIX: $S_4$ GROUP}\label{A}
The discrete group $S_{4}$ who has 24 elements is the permutation group of four objects.
The two generators \textit{S} and \textit{T} in different
irreducible presentations are given as follows
\begin{eqnarray}
  & & \textbf{1} : S = 1,  \quad  T=1 \\
  & & \textbf{1}' : S = -1, \quad  T=1 \\
  & & \textbf{2}  :
S = \left(
  \begin{array}{cc}
  0 & 1 \\
  1 & 0 \\
  \end{array}
  \right),\quad
T=\left(
  \begin{array}{cc}
  \omega & 0 \\
  0 & \omega^{2} \\
  \end{array}
  \right)\\
& & \textbf{3} :
S = \frac{1}{3}\left(
  \begin{array}{ccc}
  -1 & 2\omega & 2\omega^{2} \\
  2\omega & 2\omega^{2} & -1 \\
  2\omega^{2} & -1 & 2\omega \\
  \end{array}
  \right),\quad
T = \left(
  \begin{array}{ccc}
  1 & 0 & 0 \\
  0 & \omega^{2} & 0 \\
  0 & 0 & \omega \\
  \end{array}
  \right)\\
& & \textbf{3}' :
S = \frac{1}{3}\left(
  \begin{array}{ccc}
  1 & -2\omega & -2\omega^{2} \\
  -2\omega & -2\omega^{2} & 1 \\
  -2\omega^{2} & 1 & -2\omega \\
  \end{array}
  \right),\quad
T = \left(
  \begin{array}{ccc}
  1 & 0 & 0 \\
  0 & \omega^{2} & 0 \\
  0 & 0 & \omega \\
  \end{array}
  \right)
\end{eqnarray}
where $\omega=e^{2\pi i/3}=(i\sqrt{3}-1)/2$. In the basis we can obtain the
decomposition of the product representations and the Clebsch-Gordan factors.
The product rules of $S_{4}$ group, with $a_i$, $b_i$ as the elements of multiplet
in the product, are given by following
\begin{eqnarray}
& & \star \quad \bf{1} \otimes r = r \sim \emph{ab}_\emph{i} \\
& & \star \quad \bf{1}' \otimes 1' = 1 \sim \emph{ab} \\
& & \star \quad \bf{1}' \otimes 2 = 2  \sim \left(
    \begin{array}{c}
    ab_{1} \\
   -ab_{2} \\
    \end{array}
    \right) \\
& & \star \quad \bf{1}'\otimes 3 = 3' \sim \left(
    \begin{array}{c}
    ab_{1} \\
    ab_{2} \\
    ab_{3} \\
    \end{array}
    \right) \\
& & \star \quad \bf{1}'\otimes 3' = 3 \sim \left(
    \begin{array}{c}
    ab_{1} \\
    ab_{2} \\
    ab_{3} \\
    \end{array}
    \right)
\end{eqnarray}
The product rules with two-dimensional representation are given by:
\begin{eqnarray}
& & \star \quad \bf{2} \otimes 2 = 1 \oplus 1' \oplus 2          \nonumber\\
& & \quad \quad \textbf{1} \sim a_{1}b_{2}+a_{2}b_{1} ,
\quad \quad \textbf{1}' \sim a_{1}b_{2}-a_{2}b_{1} \nonumber\\
& & \quad \quad \textbf{2} \sim \left(
    \begin{array}{c}
    a_{2}b_{2} \\
    a_{1}b_{1} \\
    \end{array}
    \right)  \\
& & \star \quad \bf{2} \otimes 3 = 3 \oplus 3'  \nonumber\\
& & \quad \quad \textbf{3} \sim \left(
\begin{array}{c}
    a_{1}b_{2}+a_{2}b_{3} \\
    a_{1}b_{3}+a_{2}b_{1} \\
    a_{1}b_{1}+a_{2}b_{2} \\
    \end{array}
    \right) ,
\quad \quad \textbf{3}' \sim \left(
    \begin{array}{c}
    a_{1}b_{2}-a_{2}b_{3} \\
    a_{1}b_{3}-a_{2}b_{1} \\
    a_{1}b_{1}-a_{2}b_{2} \\
    \end{array}
    \right) \\
& & \star \quad \bf{2} \otimes 3' = 3 \oplus 3'  \nonumber\\
& & \quad \quad \textbf{3} \sim \left(
\begin{array}{c}
    a_{1}b_{2}-a_{2}b_{3} \\
    a_{1}b_{3}-a_{2}b_{1} \\
    a_{1}b_{1}-a_{2}b_{2} \\
    \end{array}
    \right) ,
\quad \quad \textbf{3}' \sim \left(
    \begin{array}{c}
    a_{1}b_{2}+a_{2}b_{3} \\
    a_{1}b_{3}+a_{2}b_{1} \\
    a_{1}b_{1}+a_{2}b_{2} \\
    \end{array}
    \right)
\end{eqnarray}
and the three-dimensional representations have the following product rules
\begin{eqnarray}
& & \star \quad \bf{3} \otimes 3 = 3' \otimes 3'=1 \oplus 2 \oplus 3 \oplus 3' \nonumber\\
& & \quad \quad \textbf{1} \sim a_{1}b_{1}+a_{2}b_{3}
+a_{3}b_{2} \nonumber\\
& & \quad \quad \textbf{2} \sim \left(
  \begin{array}{c}
  a_{2}b_{2}+a_{3}b_{1}+a_{1}b_{3} \\
  a_{3}b_{3}+a_{1}b_{2}+a_{2}b_{1} \\
  \end{array}
  \right)\nonumber\\
& & \quad \quad \textbf{3} \sim \left(
  \begin{array}{c}
  2a_{1}b_{1}-a_{2}b_{3}-a_{3}b_{2} \\
  2a_{3}b_{3}-a_{1}b_{2}-a_{2}b_{1} \\
  2a_{2}b_{2}-a_{3}b_{1}-a_{1}b_{3} \\
  \end{array}
  \right),
\quad \quad  \textbf{3}' \sim \left(
  \begin{array}{c}
  a_{2}b_{3}-a_{3}b_{2} \\
  a_{1}b_{2}-a_{2}b_{1} \\
  a_{3}b_{1}-a_{1}b_{3} \\
  \end{array}
  \right)
\end{eqnarray}
\begin{eqnarray}
& & \star \quad \bf{3} \otimes 3' = 1' \oplus 2 \oplus 3 \oplus 3' \nonumber\\
& & \quad \quad \textbf{1}' \sim a_{1}b_{1}+a_{2}b_{3}
+a_{3}b_{2} \nonumber\\
& & \quad \quad \textbf{2} \sim \left(
  \begin{array}{c}
  a_{2}b_{2}+a_{3}b_{1}+a_{1}b_{3} \\
  -a_{3}b_{3}-a_{1}b_{2}-a_{2}b_{1} \\
  \end{array}
  \right)   \nonumber\\
& & \quad \quad \textbf{3} \sim
    \left(
  \begin{array}{c}
  a_{2}b_{3}-a_{3}b_{2} \\
  a_{1}b_{2}-a_{2}b_{1} \\
  a_{3}b_{1}-a_{1}b_{3} \\
  \end{array}
  \right),
\quad \quad  \textbf{3}' \sim\left(
  \begin{array}{c}
  2a_{1}b_{1}-a_{2}b_{3}-a_{3}b_{2} \\
  2a_{3}b_{3}-a_{1}b_{2}-a_{2}b_{1} \\
  2a_{2}b_{2}-a_{3}b_{1}-a_{1}b_{3} \\
  \end{array}
  \right)
\end{eqnarray}
\end{appendix}

%-----------------------------------------------------------------------------------
%%%%%%%%%%%%%%%%%%%%%%%%%%%%%%%%%%%%%%%%%%%%%%%%
%%%%%%%%%%%%%%%%%%%%%%%%%%%%%%%%%%%%%%%%%%%%%%%%
\thispagestyle{empty}
%\newpage
\bibliographystyle{plain}
%\bibliography{001mybib}

\begin{thebibliography}{90}
\bibitem{Aad/2012ATLAS}
\textbf{ATLAS} Collaboration,
G.~Aad \textit{et al}., %and T.~Abajyan and B.~Abbott and others,
{\it ``Observation of a new particle in the search for the Standard Model Higgs boson
with the ATLAS detector at the LHC"},
\href{http://www.sciencedirect.com/science/article/pii/S037026931200857X}
{Phys. Lett. \textbf{B 716} (2012) 1},
[\href{https://arxiv.org/abs/1207.7214}{arXiv:1207.7214 [hep-ex]}].

\bibitem{Chatrchyan/2012CMS}
\textbf{CMS} Collaboration,
S. Chatrchyan \textit{et al}., %and V. Khachatryan and A.M. Sirunyan and others,
{\it ``Observation of a new boson at a mass of 125 GeV with the CMS experiment
at the LHC"},
\href{http://www.sciencedirect.com/science/article/pii/S0370269312008581}
{Phys. Lett. \textbf{B 716} (2012) 30},
[\href{https://arxiv.org/abs/1207.7235}{arXiv:1207.7235 [hep-ex]}].


\bibitem{Tanabashi/2018PDG}
\textbf{Particle Data Group} Collaboration, M. Tanabashi \textit{et al}.,
{\it ``Review of Particle Physics"},
\href{https://link.aps.org/doi/10.1103/PhysRevD.98.030001}
{Phys. Rev. \textbf{D 98} (2018) 030001}.
%[\href{}{}].

\bibitem{Hall/2000ANX}
L.J.~Hall, H.~Murayama and N.~Weiner,
{\it ``Neutrino Mass Anarchy"},
\href{https://link.aps.org/doi/10.1103/PhysRevLett.84.2572}
{Phys. Rev. Lett. \textbf{84} (2000) 2572},
[\href{https://arxiv.org/abs/hep-ph/9911341}{arXiv:hep-ph/9911341}].

\bibitem{Dienes/1999XDS}
K.R.~Dienes, E.~Dudas and T.~Gherghetta,
{\it ``Light neutrinos without heavy mass scales: a higher-dimensional seesaw mechanism"},
\href{http://www.sciencedirect.com/science/article/pii/S0550321399003776}
{Nucl. Phys. \textbf{B 557} (1999) 25},
[\href{https://arxiv.org/abs/hep-ph/9811428}{arXiv:hep-ph/9811428}].

\bibitem{Arkani-Hamed/2002NXD}
N.~Arkani-Hamed, S.~Dimopoulos, G.R.~Dvali and J.~March-Russell,
{\it ``Neutrino masses from large extra dimensions"},
\href{http://link.aps.org/doi/10.1103/PhysRevD.65.024032}
{Phys. Rev. \textbf{D 65} (2002) 024032},
[\href{https://arxiv.org/abs/hep-ph/9811448}{arXiv:hep-ph/9811448}].



\bibitem{Blumenhagen/2007DBR}
R.~Blumenhagen, M.~Cveti\v{c} and T.~Weigand,
{\it ``Spacetime instanton corrections in 4D string vacua:
The seesaw mechanism for D-brane models"},
\href{https://doi.org/10.1016/j.nuclphysb.2007.02.016}
{Nucl. Phys. \textbf{B 771} (2007) 113},
[\href{https://arxiv.org/abs/hep-th/0609191}{arXiv:hep-th/0609191}].

\bibitem{Ibanez/2007STN}
L.E.~Ib{\'{a}}{\~{n}}ez and A.M.~Uranga,
{\it ``Neutrino Majorana masses from string theory instanton effects"},
\href{https://doi.org/10.1088/1126-6708/2007/03/052}{JHEP \textbf{03} (2007) 052},
[\href{https://arxiv.org/abs/hep-th/0609213}{arXiv:hep-th/0609213}].

\bibitem{Antusch/2007STN}
S. Antusch, L.E.~Ib{\'{a}}{\~{n}}ez and T. Macr\`{i},
{\it ``Neutrino masses and mixings from string theory instantons"},
\href{https://doi.org/10.1088/1126-6708/2007/09/087}{JHEP \textbf{09} (2007) 087},
[\href{https://arxiv.org/abs/0706.2132}{arXiv:0706.2132 [hep-ph]}].


\bibitem{Altarelli/2010RMP}
G.~Altarelli and F.~Feruglio,
{\it ``Discrete flavor symmetries and models of neutrino mixing"},
\href{http://link.aps.org/doi/10.1103/RevModPhys.82.2701}
{Rev. Mod. Phys. \textbf{82} (2010) 2701},
[\href{https://arxiv.org/abs/1002.0211}{arXiv:1002.0211 [hep-ph]}].

\bibitem{Ishimori/2010NAD}
H.~Ishimori, T.~Kobayashi, H.~Ohki, Y.~Shimizu,
H.~Okada and M.~Tanimoto,
{\it ``Non-Abelian Discrete Symmetries in Particle Physics"},
\href{http://ptps.oxfordjournals.org/content/183/1.abstract}
{Prog. Theor. Phys. Suppl. \textbf{183} (2010) 1-163},
[\href{https://arxiv.org/abs/1003.3553}{arXiv:1003.3552 [hep-th]}].

\bibitem{King/2013RPP}
S.F.~ King and C.~Luhn,
{\it ``Neutrino mass and mixing with discrete symmetry"},
\href{http://stacks.iop.org/0034-4885/76/i=5/a=056201}
{Rep. Prog. Phys. \textbf{76} (2013) 056201},
[\href{http://www.arxiv.org/abs/1301.1340}{arXiv:1301.1340 [hep-ph]}].

\bibitem{King/2015REV}
S.F.~King,
{\it ``Models of neutrino mass, mixing and CP violation"},
\href{http://stacks.iop.org/0954-3899/42/i=12/a=123001}
{J. Phys. G \textbf{42} (2015) 123001},
[\href{http://www.arxiv.org/abs/1510.02091}{arXiv:1510.02091 [hep-ph]}].

\bibitem{King/2017UNM}
S.F.~King,
{\it ``Unified models of neutrinos, flavour and CP Violation"},
\href{http://dx.doi.org/10.1016/j.ppnp.2017.01.003}
{Prog. Part. Nucl.Phys. \textbf{94} (2017) 217},
[\href{https://arxiv.org/abs/1701.04413}{arXiv:1701.04413 [hep-ph]}].

\bibitem{Meloni/2017FMN}
D.~Meloni,
{\it ``GUT and Flavor Models for Neutrino Masses and Mixing"},
\href{https://www.frontiersin.org/article/10.3389/fphy.2017.00043}
{Frontiers in Physics \textbf{5} (2017) 43},
[\href{https://arxiv.org/abs/1709.02662}{arXiv:1709.02662 [hep-ph]}].


\bibitem{Harrison/2002HPS}
P.F.~Harrison, D.H.~Perkins and W.G.~Scott,
{\it ``Tri-bimaximal mixing and the neutrino oscillation data"},
\href{http://dx.doi.org/10.1016/S0370-2693(02)01336-9}
{Phys.~Lett.~\textbf{B 530} (2002) 167},
[\href{https://arxiv.org/abs/hep-ph/0202074}{arXiv:hep-ph/0202074}].

\bibitem{Harrison/2002TBM}
P.F.~Harrison and W.G.~Scott,
\href{http://dx.doi.org/10.1016/S0370-2693(02)01753-7}
{Phys.~Lett.~\textbf{B 535} (2002) 163},
[\href{https://arxiv.org/abs/hep-ph/0203209}{arXiv:hep-ph/0203209}].

\bibitem{Feruglio/2017NMF}
F.~Feruglio,
{\it ``Are neutrino masses modular forms?"},
\href{https://www.worldscientific.com/doi/abs/10.1142/9789813238053_0012}
{in book ``From my vast repertoire:the legacy of Guido Altarelli", S. Forte,
A.Levy and G. Ridolfi, eds},
[\href{https://arxiv.org/abs/1706.08749}{arXiv:1706.08749 [hep-ph]}].


\bibitem{Kobayashi/2018NFM}
T.~Kobayashi, K.~Tanaka, and T.~H.~Tatsuishi,
{\it ``Neutrino mixing from finite modular groups"},
\href{https://link.aps.org/doi/10.1103/PhysRevD.98.016004}
{Phys. Rev. \textbf{D 98} (2018) 016004},
[\href{https://arxiv.org/abs/1803.10391}{arXiv:1803.10391 [hep-ph]}].


\bibitem{Okada/2019MST}
H.~Okada and Y.~Orikasa,
{\it ``Modular ${S}_{3}$ symmetric radiative seesaw model"},
\href{https://link.aps.org/doi/10.1103/PhysRevD.100.115037}
{Phys. Rev. \textbf{D 100} (2019) 115037},
[\href{http://www.arxiv.org/abs/1907.04716}{arXiv:1907.04716 [hep-ph]}].

\bibitem{Kobayashi/2020SRG}
T.~Kobayashi, Y.~Shimizu, K.~Takagi, M.~Tanimoto and Takuya H. Tatsuishi,
{\it ``Modular $S_3$-invariant flavor model in SU(5) grand unified theory"},
\href{http://dx.doi.org/10.1093/ptep/ptaa055}
{Prog. Theor. Exp. Phys. \textbf{053B} (2020) 05},
[\href{http://www.arxiv.org/abs/1906.10341}{arXiv:1906.10341 [hep-ph]}].

\bibitem{Mishra/2020MLG}
S.~Mishra,
{\it ``Neutrino mixing and Leptogenesis with modular $S_3$ symmetry in the framework
of type III seesaw"},
%\href{}{ () },
[\href{https://arxiv.org/abs/2008.02095}{arXiv:2008.02095 [hep-ph]}].

%%%%%A4A4A4A4A4A4A4A4%%%%%%%%%%%%%%%%%%%%%%%%%%%%%%%%%%%%
\bibitem{Criado/2018MDN}
J.C.~Criado and F.~Feruglio,
{\it ``Modular Invariance Faces Precision Neutrino Data"},
\href{https://scipost.org/10.21468/SciPostPhys.5.5.042}
{SciPost Phys. \textbf{5} (2018) 042},
[\href{http://www.arxiv.org/abs/1807.01125}{arXiv:1807.01125 [hep-ph]}].


\bibitem{Okada/2019CPV}
H. Okada and M. Tanimoto,
{\it ``CP violation of quarks in $A_4$ modular invariance"},
%\href{}{ () },
[\href{http://arxiv.org/abs/1812.09677}{arXiv:1812.09677 [hep-ph]}].


\bibitem{Okada/2020CTA}
H. Okada and M. Tanimoto,
{\it ``Quark and lepton flavors with common modulus $\tau$ in $A_4$ modular symmetry"},
%\href{}{ () },
[\href{http://arxiv.org/abs/2005.00775}{arXiv:2005.00775 [hep-ph]}].


\bibitem{Kobayashi/2019BNV}
T.~Kobayashi, Y.~Shimizu, K.~Takagi, M.~Tanimoto, T.~H. Tatsuishi and H.~Uchida,
{\it ``Finite modular subgroups for fermion mass matrices and baryon/lepton
number violation"},
\href{https://doi.org/10.1016/j.physletb.2019.05.034}{Phys. Lett. B 794 (2019) 114},
[\href{https://arxiv.org/abs/1812.11072}{arXiv:1812.11072 [hep-ph]}].


\bibitem{Nomura/2019MAF}
T.~Nomura and H.~Okada,
{\it ``A modular $A_4$ symmetric model of dark matter and neutrino"},
\href{https://doi.org/10.1016/j.physletb.2019.134799}
{Phys. Lett. \textbf{B 797} (2019) 134799},
[\href{http://www.arxiv.org/abs/1904.03937}{arXiv:1904.03937 [hep-ph]}].


\bibitem{Nomura/2019TLP}
T.~Nomura, and H.~Okada,
{\it ``A two loop induced neutrino mass model with modular $A_4$ symmetry"},
%\href{}{ () },
[\href{http://www.arxiv.org/abs/1906.03927}{arXiv:1906.03927 [hep-ph]}].


\bibitem{Ding/2019MAF}
G.-J. Ding, and S. F. King, and X.-G. Liu,
{\it ``Modular $A_4$ symmetry models of neutrinos and charged leptons"},
\href{https://doi.org/10.1007/JHEP09(2019)074}{JHEP \textbf{09} (2019) 074},
[\href{https://arxiv.org/abs/1907.11714}{arXiv:1907.11714 [hep-ph]}].


\bibitem{Kobayashi/2020ASM}
T.~Kobayashi, Y.~Shimizu, K.~Takagi, M.~Tanimoto and Takuya H.~Tatsuishi,
{\it ``New $A_4$ lepton flavor model from $S_4$ modular symmetry"},
\href{https://doi.org/10.1007/JHEP02(2020)097}{JHEP \textbf{02} (2020) 097},
[\href{https://arxiv.org/abs/1907.09141}{arXiv:1907.09141 [hep-ph]}].


\bibitem{KingSJ/2020FMH}
Simon J. D. King, S. F. King,
{\it ``Fermion Mass Hierarchies from Modular Symmetry"},
\href{https://doi.org/10.1007/JHEP09(2020)043}{JHEP \textbf{09} (2020) 043},
[\href{http://arxiv.org/abs/2002.00969}{arXiv:2002.00969 [hep-ph]}].


\bibitem{Anda/2020GAF}
F.J. de Anda, and S. F. King, and E. Perdomo,
{\it ``$SU\mathbf{(}5\mathbf{)}$ grand unified theory with ${A}_{4}$ modular symmetry"},
\href{https://link.aps.org/doi/10.1103/PhysRevD.101.015028}
{Phys. Rev. \textbf{D 101} (2020) 015028},
[\href{https://arxiv.org/abs/1812.05620}{arXiv:1812.05620 [hep-ph]}].


\bibitem{Asaka/2020MAF}
T.~Asaka, Y.~Heo, T.~H. Tatsuishi and T.~Yoshida,
{\it ``Modular $A_4$ invariance and leptogenesis"},
\href{https://doi.org/10.1007/JHEP01(2020)144}{JHEP \textbf{01} (2020) 144},
[\href{https://arxiv.org/abs/1909.06520}{arXiv:1909.06520 [hep-ph]}].


\bibitem{Behera/2020MAL}
M.K.~Behera, S.~Mishra, S.~Singirala and R.~Mohanta,
{\it ``Implications of $A_4$ modular symmetry on Neutrino mass, Mixing and Leptogenesis
with Linear Seesaw"},
%\href{}{ () },
[\href{https://arxiv.org/abs/2007.00545}{arXiv:2007.00545 [hep-ph]}].


%%%%%%%%%%%%%%%S4S4S4S4S4S4S4S4S4S4%%%%%%%%%%%S4S4S4S4S4%%%%%%%%%%%%%%%


\bibitem{Penedo/2018LMS}
J.~Penedo and S.~Petcov,
{\it ``Lepton Masses and Mixing from Modular $S_4$ Symmetry''},
\href{http://dx.doi.org/10.1016/j.nuclphysb.2018.12.016}
{Nucl.~Phys.~\textbf{B 939} (2019) 292},
[\href{http://arxiv.org/abs/1806.11040}{arXiv:1806.11040 [hep-ph]}].

\bibitem{Novichkov/2018MSL}
P.~Novichkov, J.~Penedo, S.~Petcov, and A.~Titov,
{\it ``Modular S$_{4}$ models of lepton masses and mixing''},
\href{http://dx.doi.org/10.1007/JHEP04(2019)005}
{JHEP \textbf{04} (2019) 005},
[\href{http://arxiv.org/abs/1811.04933}{arXiv:1811.04933 [hep-ph]}].

\bibitem{MedeirosVarzielas/2019MPS}
I.~de~Medeiros~Varzielas, S.~F. King, and Y.-L. Zhou,
{\it``Multiple modular symmetries as the origin of flavor''},
\href{http://dx.doi.org/10.1103/PhysRevD.101.055033}
{Phys. Rev. \textbf{D 101} (2020) 055033},
[\href{http://arxiv.org/abs/1906.02208}{arXiv:1906.02208 [hep-ph]}].


\bibitem{Ding/2019MAS}
G.-J. Ding, S.~F. King, X.-G. Liu, and J.-N. Lu,
{\it``Modular S$_{4}$ and A$_{4}$ symmetries and their fixed points:
new predictive examples of lepton mixing''},
\href{http://dx.doi.org/10.1007/JHEP12(2019)030}
{JHEP \textbf{12} (2019) 030},
[\href{http://arxiv.org/abs/1910.03460}{arXiv:1910.03460 [hep-ph]}].

\bibitem{Kobayashi/2019ASM}
T.~Kobayashi, Y.~Shimizu, K.~Takagi, M.~Tanimoto, and T.~H. Tatsuishi,
{\it``New $A_4$ lepton flavor model from $S_4$ modular symmetry''},
\href{http://dx.doi.org/10.1007/JHEP02(2020)097}{JHEP 02 (2020) 097},
[\href{http://arxiv.org/abs/1907.09141}{arXiv:1907.09141 [hep-ph]}].

\bibitem{King/2019DMS}
S.~F. King and Y.-L. Zhou,
{\it``Trimaximal TM$_1$ mixing with two modular $S_4$ groups''},
\href{http://dx.doi.org/10.1103/PhysRevD.101.015001}
{Phys. Rev. \textbf{D 101} (2020) 015001},
[\href{http://arxiv.org/abs/1908.02770}{arXiv:1908.02770 [hep-ph]}].

\bibitem{Criado/2019LQP}
J.~C. Criado, F.~Feruglio, and S.~J. King,
{\it``Modular Invariant Models of Lepton Masses at Levels 4 and 5''},
\href{http://dx.doi.org/10.1007/JHEP02(2020)001}{JHEP \textbf{02} (2020) 001},
[\href{http://arxiv.org/abs/1908.11867}{arXiv:1908.11867 [hep-ph]}].

\bibitem{WangX/2020QMS}
X.~Wang and S.~Zhou,
{\it ``The minimal seesaw model with a modular S$_{4}$ symmetry''},
\href{http://dx.doi.org/10.1007/JHEP05(2020)017}{JHEP \textbf{05} (2020) 017},
[\href{http://arxiv.org/abs/1910.09473}{arXiv:1910.09473 [hep-ph]}].

\bibitem{WangX/2020DRS}
X.~Wang,
{\it ``A systematic study of Dirac neutrino mass models with a modular
$S_4$ symmetry''},
[\href{http://arxiv.org/abs/2007.05913}{arXiv:2007.05913 [hep-ph]}].


%%%%%%%%%%A5A5A5A5A5A5A5A5A5A5A5%%%%%%%%%%%%%%%%%%%%%%%%%%%%%%%%%%%%%%%%%%%%


\bibitem{Novichkov/2018AFM}
P.~Novichkov, J.~Penedo, S.~Petcov, and A.~Titov,
{\it ``Modular A$_{5}$ symmetry for flavour model building''},
\href{http://dx.doi.org/10.1007/JHEP04(2019)174}{JHEP \textbf{04} (2019) 174},
[\href{http://arxiv.org/abs/1812.02158}{arXiv:1812.02158 [hep-ph]}].


\bibitem{Ding/2019AFM}
G.-J. Ding, S.~F. King, and X.-G. Liu,
{\it ``Neutrino mass and mixing with $A_5$ modular symmetry''}
\href{http://dx.doi.org/10.1103/PhysRevD.100.115005}
{Phys. Rev. \textbf{D 100} (2019) 115005},
[\href{http://arxiv.org/abs/1903.12588}{arXiv:1903.12588 [hep-ph]}].


\bibitem{LiuXG/2019DCA}
X.-G. Liu and G.-J. Ding,
{\it ``Neutrino Masses and Mixing from Double Covering of Finite Modular Groups"},
%\href{}{JHEP \textbf{08} (2019) 134},
[\href{http://www.arxiv.org/abs/1907.01488}{arXiv:1907.01488 [hep-ph]}].

%%%%%DCDCDCDCDCDCDC%%%%%%%%%%%%%%%%%%%%%%%%%%%%%%%%%%%%%%%%%%%%%%%%%%%%%%%

\bibitem{LuJN/2020TXZ}
J.-N. Lu, X.-G. Liu, and G.-J. Ding,
{\it ``Modular symmetry origin of texture zeros and quark lepton unification"},
\href{http://dx.doi.org/10.1103/PhysRevD.101.115020}
{Phys. Rev. \textbf{D 101} (2020) 115020},
[\href{http://arxiv.org/abs/1912.07573}{arXiv:1912.07573 [hep-ph]}].


\bibitem{Novichkov/2020DCS}
P. P. Novichkov, J. T. Penedo, S. T. Petcov,
{\it ``Double Cover of Modular $S_4$ for Flavour Model Building"},
%\href{}{ () },
[\href{http://arxiv.org/abs/2006.03058}{arXiv:2006.03058 [hep-ph]}].


\bibitem{LiuXG/2020DCF}
X.-G. Liu, C.-Y. Yao, G.-J. Ding,
{\it ``Modular Invariant Quark and Lepton Models in Double Covering
of $S_4$ Modular Group,"},
%\href{}{ () },
[\href{http://arxiv.org/abs/2006.10722}{arXiv:2006.10722 [hep-ph]}].


\bibitem{WangX/2020DCF}
X. Wang, B. Yu, and S. Zhou,
{\it ``Double Covering of the Modular $A_5$ Group and Lepton
Flavor Mixing in the Minimal Seesaw Model"},
%\href{}{ () },
[\href{http://arxiv.org/abs/2010.10159}{arXiv:2010.10159 [hep-ph]}].


\bibitem{YaoCY/2020DCF}
C.-Y. Yao, X.-G. Liu, G.-J. Ding
{\it ``Fermion Masses and Mixing from Double Cover and Metaplectic Cover
of $A_5$ Modular Group"},
%\href{}{ () },
[\href{http://arxiv.org/abs/2011.03501}{arXiv:2011.03501 [hep-ph]}].


%%%%%%%%%%%%%%%%%%%%%%%%%%%%%%%%%%%%%%%%%%%%%%%%%%%%%%%%%%%%%%%%%%%%%%%%%%%%%%%

\bibitem{Perez/2007SAS}
P.~Fileviez P\'{e}rez,
{\it ``Supersymmetric adjoint SU(5)"},
\href{http://dx.doi.org/10.1103/PhysRevD.76.071701}
{Phys. Rev. \textbf{D 76} (2007) 071701},
[\href{http://arxiv.org/abs/0705.3589}{arXiv:0705.3589 [hep-ph]}].

\bibitem{Minkowski/1977SS}
P.~Minkowski,
{\it ``$\mu\rightarrow e\gamma$ at a rate of one out of $10^9$ muon decays?"},
\href{http://www.sciencedirect.com/science/article/pii/037026937790435X}
{Phys.~Lett.~\textbf{B 67} (1977) 421}.

\bibitem{Yanagida/1979SS}
T.~Yanagida, 1979,
{\it ``Horizontal symmetry and masses of neutrinos"},
in the proceedings of the
{\it ``Workshop on Unified Theory and Baryon Number in the Universe"},
O. Sawada and A. Sugamoto eds., KEK, Tsukuba, Japan (1979).

\bibitem{Mohapatra/1980SS}
R.N.~Mohapatra and G.~Senjanovi\'{c},
{\it ``Neutrino Mass and Spontaneous Parity Nonconservation"},
\href{http://link.aps.org/doi/10.1103/PhysRevLett.44.912}
{Phys.~Rev.~Lett.~ \textbf{44} (1980) 912}.

\bibitem{Gell-Mann/1979sug}
M.~Gell-Mann, P.~Ramond and R.~Slansky,
{\it ``Complex Spinors and Unified Theories"}, In {\it``Supergravity"},
D.Z.~Freedman and F.~van Nieuwenhuizen eds., North Holland,
Amsterdam, The Netherlands (1979),
[\href{https://arxiv.org/abs/1306.4669}{arXiv:1306.4669 [hep-th]}].

\bibitem{Lazarides/1981SGT}
G. Lazarides and Q. Shafi and C. Wetterich,
{\it ``Proton lifetime and fermion masses in an SO(10) model"},
\href{https://doi.org/10.1016/0550-3213(81)90354-0}
{Nucl. Phys. \textbf{B 181} (1981) 287}.
%[\href{}{}].

\bibitem{Schechter/1980NUM}
J.~Schechter and J.W.F.~Valle,
{\it ``Neutrino masses in SU(2)\ensuremath{\bigotimes}U(1) theories"},
\href{http://link.aps.org/doi/10.1103/PhysRevD.22.2227}
{Phys. Rev. \textbf{D 22} (1980) 2227}.
%[\href{}{}].

\bibitem{Mohapatra/1981NMV}
R.N.~Mohapatra and S.~Goran,
{\it ``Neutrino masses and mixings in gauge models with spontaneous parity violation"},
\href{http://link.aps.org/doi/10.1103/PhysRevD.23.165}
{Phys. Rev. \textbf{D 23} (1981) 165}.
%[\href{}{}].

\bibitem{Foot/1989SS3}
R. Foot, H. Lew, X.-G. He and G.C.~Joshi,
{\it ``See-saw neutrino masses induced by a triplet of leptons"},
\href{http://dx.doi.org/10.1007/BF01415558}{Z. Phys. \textbf{C 44} (1989) 441}.
%[\href{}{}].


\bibitem{Perez/2016RAD}
P.~Fileviez P\'{e}rez,
{\it ``Renormalizable adjoint SU(5)"},
\href{http://dx.doi.org/10.1016/j.physletb.2007.07.075}
{Phys. Lett. \textbf{B 654} (2007) 189},
[\href{https://arxiv.org/abs/hep-ph/0702287}{arXiv:hep-ph/0702287}].


\bibitem{Binosi/2004JXD}
D. Binosi and L. Theu{\ss}l,
{\it ``JaxoDraw: A graphical user interface for drawing Feynman diagrams"},
\href{https://doi.org/10.1016/j.cpc.2004.05.001}
{Comput. Phys. Commun. \textbf{161} (2004) 76},
[\href{https://arxiv.org/abs/hep-ph/0309015}{arXiv:hep-ph/0309015}].

\bibitem{Binosi/2009JXD}
D. Binosi, J. Collins, C. Kaufhold and L. Theu{\ss}l,
{\it ``JaxoDraw: A graphical user interface for drawing Feynman diagrams.
Version 2.0 release notes"},
\href{http://dx.doi.org/10.1016/j.cpc.2009.02.020}
{Comput. Phys. Commun. \textbf{180} (2009) 1709},
[\href{https://arxiv.org/abs/0811.4113}{arXiv:0811.4113 [hep-ph]}].


\bibitem{Antusch/2013RUN}
S.~Antusch, V.~Maurer,
{\it ``Running quark and lepton parameters at various scales"},
\href{http://dx.doi.org/10.1007/JHEP11(2013)115}{JHEP \textbf{11} (2013) 115},
[\href{https://arxiv.org/abd/hep-ph/1306.6879}{arXiv:1306.6879 [hep-ph]}].


\bibitem{Georgi/1979GJ}
H.~Georgi and C.~Jarlskog,
{\it ``A new lepton-quark mass relation in a unified theory"},
\href{http://dx.doi.org/10.1016/0370-2693(79)90842-6}
{Phys. Lett. \textbf{B 86} (1979) 297}.


\bibitem{Antusch/2013GUT}
S.~Antusch, C.~Gross, V.~Maurer and C.~Sluka,
{\it ``A flavour GUT model with $\theta^{PMNS}_{13}\simeq\theta_C/\sqrt{2}$"},
\href{http://dx.doi.org/10.1016/j.nuclphysb.2013.11.003}
{Nucl. Phys. \textbf{B 877} (2013) 772},
[\href{https://arxiv.org/abs/1305.6612}{arXiv:1305.6612 [hep-ph]}].

\bibitem{Gehrlein/2015GRF}
J.~Gehrlein, J.P. Oppermann, D.~Sch\"{a}fer and M.~Spinrath,
{\it ``An SU(5)$\times A_{5}$ golden ratio flavour model"},
\href{http://dx.doi.org/10.1016/j.nuclphysb.2014.11.023}
{Nucl. Phys. \textbf{B 890} (2015) 539},
[\href{http://www.arxiv.org/abs/1410.2057}{arXiv:1410.2057 [hep-ph]}].

\bibitem{Bjorkeroth/2015ASU}
F.~Bj{\"o}rkeroth, F.J. de Anda, I.~de Medeiros Varzielas and S.F.~King,
{\it ``Towards a complete A$_{4}\times$SU(5) SUSY GUT"},
\href{http://dx.doi.org/10.1007/JHEP06(2015)141}{JHEP \textbf{06} (2015) 141},
[\href{https://arxiv.org/abs/1503.03306}{arXiv:1503.03306 [hep-ph]}].

\bibitem{Zhao/2016GUT}
Y.~Zhao and P.-F.~Zhang,
{\it ``SUSY SU(5) $\times S_{4}$ GUT Flavor Model for Fermion Masses and Mixings
with Adjoint, Large $\theta^{PMNS}_{13}$"},
\href{https://doi.org/10.1007/JHEP06(2016)032}
{JHEP \textbf{06} (2016) 032},
[\href{https://arxiv.org/abs/1402.5834}{arXiv:1402.5834 [hep-ph]}].

\bibitem{Antusch/2014GPN}
S.~Antusch, S.F. King and M.~Spinrath,
{\it ``GUT predictions for quark-lepton Yukawa coupling ratios with messenger masses
from non-singlets"},
\href{http://link.aps.org/doi/10.1103/PhysRevD.89.055027}
{Phys. Rev. \textbf{D 89} (2014) 055027},
[\href{https://arxiv.org/abs/1311.0877}{arXiv:1311.0877 [hep-ph]}].


\bibitem{Hall/1994THS}
L.J.~Hall, R.~Rattazzi and U.~Sarid,
{\it ``Top quark mass in supersymmetric SO(10) unification"},
\href{https://link.aps.org/doi/10.1103/PhysRevD.50.7048}
{Phys. Rev. \textbf{D 50} (1994) 7048},
[\href{https://arxiv.org/abs/hep-ph/9306309}{arXiv:hep-ph/9306309}].


\bibitem{Carena/1994YTH}
M. Carena, M. Olechowski, S. Pokorski and C.E.M. Wagner,
{\it ``Electroweak symmetry breaking and bottom-top Yukawa unification"},
\href{https://doi.org/10.1016/0550-3213(94)90313-1}
{Nucl. Phys. \textbf{B 426} (1994) 269},
[\href{https://arxiv.org/abs/hep-ph/9402253}{arXiv:hep-ph/9402253}].


\bibitem{Hempfling/1994UTH}
R.~Hempfling,
{\it ``Yukawa coupling unification with supersymmetric threshold corrections"},
\href{https://link.aps.org/doi/10.1103/PhysRevD.49.6168}
{Phys. Rev. \textbf{D 49} (1994) 6168}.
%[\href{}{}].


\bibitem{Blazek/1995FTH}
T.~Bla\v{z}ek, S.~Raby and S.~Pokorski,
{\it ``Finite supersymmetric threshold corrections to CKM matrix elements
in the large tan\ensuremath{\beta} regime"},
\href{https://link.aps.org/doi/10.1103/PhysRevD.52.4151}
{Phys. Rev. \textbf{D 52} (1995) 4151},
[\href{https://arxiv.org/abs/hep-ph/9504364}{arXiv:hep-ph/9504364}].


\bibitem{Antusch/2018CSD}
S.~Antusch, C.~Hohl, C.~K.~Khosa and V.~Susi$\check{\textrm{c}}$,
{\it ``Predicting $\delta^{PMNS}$, $\theta^{PMNS}_{23}$ and fermion mass ratios
from flavour GUTs with CSD2"},
\href{https://doi.org/10.1007/JHEP12(2018)025}{JHEP \textbf{12} (2018) 025},
[\href{https://arxiv.org/abs/1808.09364}{arXiv:1808.09364 [hep-ph]}].

\bibitem{Antusch/2009NGUT}
S.~Antusch and M.~Spinrath,
{\it ``New GUT predictions for quark and lepton mass ratios confronted
with phenomenology"},
\href{http://link.aps.org/doi/10.1103/PhysRevD.79.095004}
{Phys. Rev. \textbf{D 79} (2009) 095004},
[\href{https://arxiv.org/abs/0902.4644}{arXiv:0902.4644 [hep-ph]}].



\bibitem{Esteban/2020GNF}
I.~Esteban, M.~C.~Gonzalez-Garcia, M.~Maltoni, T.~Schwetz and A.~Zhou,
{\it ``The fate of hints: updated global analysis of three-flavor neutrino oscillations"},
\href{http://dx.doi.org/10.1007/JHEP09(2020)178}{JHEP \textbf{09} (2020) 178}
[\href{https://arxiv.org/abs/2007.14792}{arXiv:2007.14792 [hep-ph]}].


\bibitem{PlanckCollaboration/2020CVI}
\textbf{PLANCK} Collaboration, N. Aghanim \textit{et al}.,
{\it ``Planck 2018 results. VI. Cosmological parameters"},
\href{https://doi.org/10.1051/0004-6361/201833910}{A\&A \textbf{641} (2020) A6},
[\href{https://arxiv.org/abs/1807.06209}{arXiv:1807.06209 [astro-ph.CO]}].


\bibitem{Gando/2016KLZ}
\textbf{KamLAND-ZEN} Collaboration, A. Gando \textit{et al}.,
{\it ``Search for Majorana Neutrinos near the Inverted Mass Hierarchy Region
with KamLAND-Zen"},
\href{http://dx.doi.org/10.1103/PhysRevLett.117.082503}
{Phys. Rev. Lett.~\textbf{117} (2016) 082503},
[\href{http://www.arxiv.org/abs/1605.02889}{arXiv:1605.02889 [hep-ex]}].


\bibitem{Aker/2019KTR}
\textbf{KATRIN} Collaboration, M. Aker \textit{et al}.,
{\it ``Improved Upper Limit on the Neutrino Mass from a Direct Kinematic
Method by KATRIN"},
\href{https://link.aps.org/doi/10.1103/PhysRevLett.123.221802}
{Phys. Rev. Lett.~\textbf{123} (2019) 221802},
[\href{http://www.arxiv.org/abs/1909.06048}{arXiv:1909.06048 [hep-ex]}].


\bibitem{Vagnozzi/2017NMS}
S.~Vagnozzi, E.~Giusarma, O.~Mena, K.~Freese, M.~Gerbino, S.~Ho, M.~Lattanzi,
{\it ``Unveiling $\ensuremath{\nu}$ secrets with cosmological data:
Neutrino masses and mass hierarchy"},
\href{https://link.aps.org/doi/10.1103/PhysRevD.96.123503}
{Phys. Rev. \textbf{D 96} (2017) 123503},
[\href{https://arxiv.org/abs/1701.08172}{arXiv:1701.08172 [astro-ph.CO]}].


\bibitem{Antusch/2005MPT}
S.~Antusch, J.~Kersten, M.~Lindner, M.~Ratz and M.A.~Schmidt,
{\it ``Running neutrino mass parameters in seesaw scenarios"},
\href{http://iopscience.iop.org/1126-6708/2005/03/024}
{JHEP \textbf{03} (2005) 024},
[\href{https://arxiv.org/abs/0501272}{arXiv:0501272 [hep-ph]}].

\bibitem{Blanchet/2008BVL}
S.~Blanchet and Pavel Fileviez P{\'{e}}rez,
{\it ``Baryogenesis via leptogenesis in adjoint SU(5)"},
\href{https://doi.org/10.1088/1475-7516/2008/08/037}
{JCAP \textbf{08} (2008) 037},
[\href{https://arxiv.org/abs/0807.3740}{arXiv:0807.3740 [hep-ph]}].

\bibitem{Kannike/2011NLG}
K.~Kannike and Dmitry V.Zhuridov,
{\it ``New solution for neutrino masses and leptogenesis in adjoint SU(5)"},
\href{https://doi.org/10.1007/JHEP07(2011)102}{JHEP \textbf{07} (2011) 102},
[\href{https://arxiv.org/abs/1105.4546}{arXiv:1105.4546 [hep-ph]}].

%\bibitem{}
%,
%%{\it ``"},
%\href{}{ () },
%[\href{}{}].

%\bibitem{}
%,
%%{\it ``"},
%\href{}{ () },
%[\href{}{}].

%\bibitem{}
%,
%%{\it ``"},
%\href{}{ () },
%[\href{}{}].

\end{thebibliography}

\end{document}